\documentclass[aps,pre,reprint,superscriptaddress]{revtex4-1}

\usepackage{multirow}
\usepackage{graphicx}
\usepackage{dcolumn}
\usepackage{bm}
\usepackage{placeins}
\usepackage{color}

\begin{document}
\preprint{}

\title{Systems-level approach to uncovering diffusive states and their transitions from single particle trajectories}
\author{Peter K. Koo}
\affiliation{Department of Physics, Yale University, New Haven, CT}
\affiliation{Integrated Graduate Program in Physical and Engineering Biology,
Yale University, New Haven, Connecticut 06520}
\author{Simon G. J. Mochrie}
\affiliation{Department of Physics, Yale University, New Haven, CT}
\affiliation{Integrated Graduate Program in Physical and Engineering Biology,
Yale University, New Haven, Connecticut 06520}
\affiliation{Department of Applied Physics, Yale University, New Haven, CT}

\email{simon.mochrie@yale.edu}

\date{\today}

\begin{abstract}
The stochastic motions of a diffusing particle contain information concerning the particle's interactions
 with binding partners and with its local environment.
 However, accurate determination of the underlying diffusive properties, beyond normal diffusion, has remained challenging
 when analyzing particle trajectories on an individual basis.
 Here, we introduce the maximum likelihood estimator (MLE) for confined diffusion and fractional Brownian motion.
 We demonstrate that this MLE yields improved estimation over traditional mean square displacement analyses.
 We also introduce  a model selection scheme (that we call mleBIC) that classifies individual
 trajectories to a given diffusion mode.
 We demonstrate the statistical limitations of classification via mleBIC using simulated data. 
 To overcome these limitations,
 we introduce a new version of perturbation expectation-maximization (pEMv2),
 which simultaneously analyzes a collection of particle trajectories to uncover
  the system of interactions which give rise to unique normal and/or non-normal diffusive states
  within the population.
  We test and evaluate the performance of pEMv2 on various sets of simulated particle trajectories,
   which transition among several modes of normal and non-normal diffusion,
     highlighting the key considerations for employing this analysis methodology.
 \end{abstract}

\pacs{}

\maketitle


\section{INTRODUCTION}

Single particle tracking  (SPT) offers the ability to non-invasively probe
at sub-diffraction-limit resolution the spatio-temporal motions of individual fluorescently-labelled proteins (FPs)
 inside living cells.
Because the different interactions, that a FP undergoes inside a cell,
give rise to different types of diffusive motion,  SPT data encode each FP's interactions
with other particles and with its local envrionment:
Biochemical binding interactions can lead to different diffusivities 
if the FP can bind to different substrates \cite{persson2013extracting};
interactions with the cellular medium can give rise to
 anomalous diffusion \cite{weber2012analytical, valentine2001investigating,parry2014bacterial} or can lead to confined motions  \cite{kusumi1993confined,rossier2012integrins}.
Thus,  important goals of SPT measurements are
(1) to infer these interactions from an analysis of  protein trajectories 
%
and  (2) to determine the spatio-temporal kinetics of each interaction.

To uncover this information,
the number of unique diffusive states, as well as each such state's diffusion mode and its diffusion properties,
must be inferred from the proteins' trajectories,
 along with the ability to classify which portions of each trajectory correspond to a given diffusive state,
 thus allowing for the determination of the underlying transition kinetics
 and the spatio-temporal locations of particular diffusive states and their transitions within the cell.  
 
Previous  work 
\cite{saxton1997single2, savin2005static, monnier2015inferring, ott2013single, das2009hidden, yang2008detection, duits2009mapping, slator2015detection, apgar2000multiple, tseng2002micromechanical}
that seeks to assess dynamic heterogeneity in tracking data has been reviewed by us
 in Ref.~\cite{Koo}.
The traditional approach for analyzing  the diffusive properties of individual particle trajectories
is by fitting each trajectory's time-averaged mean square displacement (taMSD)
to a corresponding diffusion model \cite{saxton1997single}.
However,  the way in which the taMSD is usually calculated
results in an statistically-complex representation of the underlying diffusion process,
especially for short trajectories (Supplemental Materials), rendering the taMSD unreliable.
Thus, an unweighted least squares regression
against the taMSD yields statistically inefficient estimation of the diffusion model parameters.
Improved estimation can be achieved by analyzing longer trajectories, albeit the same interaction
must persist throughout the duration of the trajectory, which is an
increasingly unlikely condition in the complex environment inside living cells.
Alternatively, ensemble-averaging  taMSD curves across particle trajectories,
 which share the same underlying diffusive properties,
is another route for bolstering the statistics and thus better representing the underlying diffusive behavior.
However, because the diffusive properties of each trajectory are not known \textit{a priori},
how to sort trajectories into groups that share diffusive properties, and therefore may be averaged together,
 is not straightforward.
 
Because of the drawbacks of taMSD analysis,
a number of
alternatives have emerged
for determining diffusion parameters,
namely the maximum likelihood estimator (MLE) \cite{berglund2010statistics}, optimal least squares fitting (OLSF) \cite{michalet2010mean}, and the covariance-based estimator (CVE)  \cite{vestergaard2014optimal}.
These approaches  have  demonstrated improved estimation in comparison with traditional taMSD analysis.
Importantly, however, to-date these approaches, which do properly account for localization noise sources, have only been shown to be applicable to particle trajectories undergoing normal diffusion. 

Recently, systems-level analyses, namely variational Bayes single particle tracking (vbSPT) \cite{persson2013extracting}
and perturbation expectation-maximization (pEM) \cite{Koo}, have  demonstrated that the limited statistics of individual particle trajectories can be augmented by simultaneously analyzing a population of particle trajectories to uncover the number of unique diffusive states and their corresponding diffusive properties.
However, both of these methods have their own limitations.
While vbSPT allows for transitions between different diffusive states, 
it fails to properly account for experimental noise sources, compromising vbSPT's ability to reliably extract the correct number of diffusive states and each state's
diffusive properties in some situations \cite{Koo}. 
On the other hand, while pEM properly accounts for experimental noise sources, it assumes that diffusive properties are constant throughout the duration of each trajectory.
Thus, pEM is only suitable to analyze particle tracks sampled at sufficiently short
timescales that transitions between different diffusive states may be neglected.
In addition, both methods make a short-time diffusion approximation, thereby
effectively assuming that every particle trajectory undergoes normal diffusion.
In fact, however, diffusing proteins interact with the complex environment in living cells,
which can lead to non-normal diffusive behavior, including, for example,
 confined diffusion within focal adhesions \cite{rossier2012integrins}
 and membrane corals \cite{kusumi1993confined},
 in which a labelled protein is tethered to a particular fixed location within a cell, 
and sub-diffusive behavior in the bacterial cytoplasm 
\cite{weber2012analytical, valentine2001investigating,parry2014bacterial},
which may be the result of the complex viscoelastic properties
 of this medium
  \cite{weber2012analytical}.
 
Thus, the short-time diffusion approximation made by pEM and vbSPT
does not necessarily hold on experimentally relevant time scales.

In the present paper, we present an overall methodology, comprising a number of advances, that overcome these limitations: First, we extend Berglund's MLE framework to determine the diffusion parameters for canonical modes of non-normal  diffusion, namely confined diffusion and fractional Brownian motion (fBm); Second, we introduce a model selection scheme, 
that we term mleBIC, which classifies individual trajectories to a given diffusion model; Third, we extend the pEM framework to be able to uncover non-normal diffusion modes and transitions between different diffusive states within particle trajectories.  We also give empirical guidelines for the sort of data likely to be necessary to successfully apply our methodology.

Specifically, in Sec. \ref{sec:ML}, we demonstrate the improved performance of MLE against traditional taMSD analysis
 on various sets of simulated particle trajectories undergoing non-normal diffusion across a wide parameter spectrum.  
Since the diffusion mode of each experimental particle track is not known \textit{a priori},
in  Sec. \ref{mleBIC}, 
we introduce a model selection scheme, based on the Bayesian information criterion (BIC), 
that we call  mleBIC,
for classifying individual trajectories to a given diffusion model.
By applying mleBIC to synthetic trajectories  undergoing various modes of non-normal diffusion,
both without and with localization noise, we illustrate, by example, the statistical limits of mleBIC's classification.
In general, we find that, even though MLE estimation is quite reliable for determining diffusion parameters,
classification to determine the correct underlying diffusion model depends strongly on the length of the trajectory,
and is only accurate for sufficiently long trajectories.
Moreover, resolving the level of heterogeneity within a population of trajectories, that realize different diffusion modes,
remains challenging.
Consequently, the SPT analysis goals defined above -- specifically, uncovering the number of diffusive states and their properties and
transitions --
cannot generally and reliably be achieved from an analysis that treats individual particle trajectories independently.

Therefore, in Section \ref{pEM},  we turn to a systems-level analysis:
We present a major extension of the pEM framework, that we call pEM version 2 (pEMv2),
that  seeks to uncover the system of diffusive behaviors arising from distinct physical interactions by: (1) identifying the number of unique diffusive states (normal or non-normal diffusion modes), (2) determining the diffusive properties of each diffusive state, and (3) classifying individual trajectories to particular diffusive states
to reveal the spatio-temporal dynamics of each diffusive behavior in reference to the cell.
In addition to now being applicable to non-normal modes of diffusion,
importantly, pEMv2 eases the other important constraint on pEM,
namely that the diffusive state remain the same throughout the trajectory.
It accomplishes this by splitting long trajectories into equally-sized bins of smaller trajectories, thus enabling
transitions between different diffusive states to be accounted for.
We test the performance of pEMv2 on various sets of synthetic particle trajectories to gain
better intuition concerning its capabilities and limitations in reference to the free parameters in the analysis.
We show that in many case pEMv2 is indeed
able to uncover and characterize normal/non-normal diffusion modes and the transitions between them.
Thus, pEMv2 represents a powerful new analysis tool  for accurately characterizing the interactions of diffusing proteins in live cells,
and it brings us a major step closer to being
able to understand  spatio-temporal biochemistry inside living cells via SPT.

\section{Results and Discussion}

\begin{table*}[!htbp]
\footnotesize
\centering{
\begin{tabular}{|c|c|}
\hline 
Mode & Covariance matrix ($\mu\rm{m}^2$)\\
\hline & \\ 
Normal  &  $
\mathbf{\Sigma}^{normal} (i,j)
= \left\{  
        \begin{array}{ll}
             2D\Delta t + 2\sigma^2 - \frac{2}{3}D\Delta t & \quad ,~j=i \\
             -\sigma^2 + \frac{1}{3}D\Delta t & \quad ,~j= i\pm 1\\
             0 & \quad ,~{\rm otherwise}
        \end{array}       
        \right.
$\\ [2ex]
& \\
\hline & \\
Confined & $
\mathbf{\Sigma}^{confined}(i,j) 
	 = \mathbf{\tilde{\Sigma}}^{confined}(i,j) + \left\{  
        \begin{array}{ll}
         2\sigma^2 - \frac{1}{6} \left( 2\mathbf{\tilde{\Sigma}}^{confined}(i,i) -  2\mathbf{\tilde{\Sigma}}^{confined}(i,i+1)\right) & \quad ,~j=i \\
           - \sigma^2 - \frac{1}{6} \left( 2\mathbf{\tilde{\Sigma}}^{confined}(i,j) -  \mathbf{\tilde{\Sigma}}^{confined} (i,j-1)- \mathbf{\tilde{\Sigma}}^{confined}({i,j+1})\right)  & \quad ,~ j=i \pm 1 \\
           -  \frac{1}{6} \left( 2\mathbf{\tilde{\Sigma}}^{confined}(i,j) -  \mathbf{\tilde{\Sigma}}^{confined}({i,j-1}) - \mathbf{\tilde{\Sigma}}^{confined}(i,j+1)\right)  & \quad ,~  {\rm otherwise}
        \end{array}       
        \right.
$\\ [2ex] &\\
& where \\
& $
\mathbf{\tilde{\Sigma}}^{confined} (i,j)
	 = \left\{  
        \begin{array}{ll}
             \frac{L^2}{6} - \frac{16L^2}{\pi^4} \sum_{k=1,odd}^\infty \frac{1}{k^4} \Phi(1)  & \quad ,~j=i \\
             \frac{-L^2}{12} + \frac{8L^2}{\pi^4} \sum_{k=1,odd}^\infty \frac{1}{k^4}  \Phi(1) \left(2-\Phi(1) \right) & \quad ,~j= i\pm 1\\
             \frac{8}{\pi^4} \sum_{k=1,odd}^\infty \frac{1}{k^4}\left( -2\Phi({j-i+1}) + \Phi({j-i})+ \Phi({j-i+2}) \right) & \quad ,~{\rm otherwise}
        \end{array}       
        \right.
$\\ 
& where $\Phi(n) = \exp\left[-\left( \frac{k\pi}{L} \right)^2 D n \Delta t \right]$.\\ [2ex]
& \\
\hline & \\
fBm \cite{backlund2015chromosomal}
& $\mathbf{\Sigma}^{fBm} (i,j)
	 = \left\{  
        \begin{array}{ll}
        \frac{2D\Delta t^\alpha}{(\alpha+2)(\alpha+1)}\left(A(1) - 2\right) + 2\sigma^2 & \quad ,~j=i\\
         \frac{D\Delta t^\alpha}{(\alpha+2)(\alpha+1)}\left(A(2) - 2A(1) + 2\right) - \sigma^2; & \quad ,~j= i\pm 1\\
         \frac{D\Delta t^\alpha}{(\alpha+2)(\alpha+1)} \left( A(\left| j-i+1\right|) - 2A(\left| j-i \right|) + A(\left| j-i-1\right|)\right) & \quad ,~{\rm otherwise}
        \end{array}       
        \right.
$\\ [2ex] &\\
& where $A(n) = (n+1)^{\alpha+2} + (n-1)^{\alpha+2} - 2n^{\alpha+2}$
\\ [2ex]
\hline & \\ 
Immobile &  $
\mathbf{\Sigma}^{immobile} (i,j)
= \left\{  
        \begin{array}{ll}
             2\sigma^2 & \quad ,~j=i \\
             -\sigma^2 & \quad ,~j= i\pm 1\\
             0 & \quad ,~{\rm otherwise}
        \end{array}       
        \right.
$\\ [2ex]
& \\
\hline
\end{tabular}
}
\caption{Analytical covariance matrix of particle track displacements separated in time by $\Delta t$ for canonical diffusion models, namely normal diffusion, confined diffusion, fractional Brownian motion, and an immobile model  \emph{with} localization noise corrections,
assuming that the camera exposure time equals the frame duration, $\Delta t$.
$D$ is the diffusion coefficient, $L$ is the confinement size for confined diffusion,
and $\alpha$ is the anomalous exponent for fBm.\\
}
\label{table:covariancenoise}
\end{table*}

\subsection{\label{sec:ML}Maximum likelihood framework}

The one-dimensional (1D) stochastic increments of a diffusing particle undergoing a stationary Gaussian process are given according to \cite{gillespie1996mathematics}: 
\begin{eqnarray}
\label{eqn:stochastic1}
	x({i+1}) = x(i) + \mathbf{\Sigma}({i,j})^{1/2} W(j),
\label{EQ1}
\end{eqnarray}
where $x(i)$ is the $x$-coordinate of the particle's position at time step $i$,
$W(j)$ is a standard Brownian motion with the properties:
$\left< W(j)\right> = 0$ and $\left< W(i), W(j)\right> = \delta_{i,j}$, where
$\delta_{i,j}$ is the Kronecker delta, $\mathbf{\Sigma}({i,j})$ is the
covariance matrix of the particle's $x$-displacements
at time steps $i$ and $j$. 
Eq.~\ref{eqn:stochastic1} employs the Einstein summation convention in which a sum over
$j$ is implied.

It follows from  Eq. \ref{eqn:stochastic1}
that the likelihood function, $P(\Delta \mathbf{x}|\mathbf{\Sigma})$, is given by a multivariate Gaussian distribution according to:
\begin{eqnarray}
 \label{eqn:likelihood}
P(\Delta \mathbf{x}|\mathbf{\Sigma}) &=& \frac{1}{( 2 \pi)^{N/2} | \mathbf{\Sigma}|^{1/2}}  \exp{\left[-\frac{1}{2}  \Delta \mathbf{x}^T \mathbf{\Sigma}^{-1} \Delta \mathbf{x} \right]}, 
 \end{eqnarray}
where $\Delta \mathbf{x}$ represents the vector of the $N$ particle track displacements,
$\{ \Delta \mathbf{x}(n) \}_{n=1}^{N}$, and
$\Delta \mathbf{x}^T$ is
its transpose.
$|\mathbf{\Sigma}|$ is the determinant of the covariance matrix, and $\mathbf{\Sigma}^{-1}$ is its inverse.
Eq. \ref{eqn:likelihood} is the likelihood function that we seek to maximize.
The dependence of the covariance matrix  in EQ.~\ref{eqn:likelihood}
on model parameters for several canonical modes of diffusion is
give in Table~\ref{table:covariancenoise}.

For normal diffusion, the presence of experimental noise sources, namely static localization noise, which is the uncertainty due to a finite number of photons emitted from a fluorophore during a camera's exposure time, and dynamic localization noise, which is the uncertainty caused by the motions of the fluorophore during a camera's exposure time,
has been shown to contribute nearest-neighbor covariance terms \cite{berglund2010statistics}. 
In the Supplemental Materials, these calculations are extended to incorporate static localization noise
into the covariance terms for
non-normal diffusion
with the result that 
\begin{eqnarray}
\label{eqn:covariance}
	{\bf \Sigma}^{static} (i,j)
	 = \left\{
        \begin{array}{ll}
             2 \sigma ^2 & \quad j=i \\
            -\sigma ^2 & \quad j= i\pm 1\\
             0 & \quad {\rm otherwise}.
        \end{array}
        \right.
\end{eqnarray}

Assuming that the camera exposure time equals $\Delta t$, which is the usual situation in SPT measurements,
the dynamic localization noise contribution
to the covariance matrix
for
normal diffusion and confined diffusion,
 may  be shown to be given approximately  by:
\begin{eqnarray}
\label{EQ4}
\mathbf{\Sigma}^{dynamic}& & (i,j) \approx \\ \nonumber
- \frac{1}{6} & & \left( 2\mathbf{\tilde{\Sigma}}(i,j)  -  \mathbf{\tilde{\Sigma}}(i+1,j) - \mathbf{\tilde{\Sigma}}(i,j+1)\right).
\end{eqnarray}
A derivation of Eq.~\ref{EQ4} is given in the Supplemental Materials.
For fBm, the contribution of dynamic localization noise to the covariance matrix 
is derived in Ref. \cite{backlund2015chromosomal}.
 As also shown in the Supplemental Materials,
corrections for static localization noise,
$\mathbf{\Sigma}^{static}$, and dynamic localization noise,
$\mathbf{\Sigma}^{dynamic}$,
contribute additively to the covariance matrix:
\begin{equation}
\mathbf{\Sigma} = \mathbf{\tilde{\Sigma}} + \mathbf{\Sigma}^{static} + \mathbf{\Sigma}^{dynamic},
\end{equation}
where $\mathbf{\tilde{\Sigma}}$
is the covariance matrix  in the absence of noise  (Appendix \ref{covariance}).
Analytical results for the covariance matrix, incorporating
localization noise corrections, for three canonical modes of diffusion,
including an immobile particle model, are given in Table~\ref{table:covariancenoise}.
The likelihood function is maximized numerically as described in Sec. \ref{mleanalysis}.

To validate the performance of our maximum likelihood framework,
we generated various sets of synthetic particle trajectories,  corresponding to
different modes of diffusion, as described in the Methods (Sec. \ref{methods}).
For confined diffusion, trajectories were simulated with a number of confinement sizes from
$0.25$ to $5$~$\mu\rm{m}$;
for fBm,  trajectories were simulated with a number of anomalous exponents from $0.25$ to $1.75$.
For normal and confined diffusion, the trajectories were simulated with a diffusion coefficient of
$D_{sim} = 0.3$~$\mu\rm{m}^2\rm{s}^{-1}$.  
For fBm,
 the trajectories were simulated with a ``diffusion coefficient'' of
$D_{sim} = 0.3$~$\mu\rm{m}^2\rm{s}^{-\alpha}$.  
Dynamic localization noise was added
by first simulating particle positions separated by
 ``micro'' time steps of
$\delta t = \Delta t/32$,
and then by averaging blocks of 32 of these positions together
to produce positions separated by time steps of $\Delta t = 32$~ms.
The net effect is to mimic experimental motion-blurred positions, corresponding
to a camera exposure time equal to the frame duration of $\Delta t = 32$~ms.
Static localization noise was included by adding a normally distributed random number
with zero mean and variance, $\sigma_{sim}^2$, to each motion-blurred position,
where
 $\sigma_{sim} = 0.04$ $\mu$m
 (Methods (Sec. \ref{methods})).
 For each set of diffusion parameters, we generated sets of particle trajectories with  track lengths
 $N= \{30, ~60, ~120, ~240\}$ steps.
 To maintain the same level of positional information across all sets of synthetic particle trajectories,
 the total number of particle positions across each simulation set was constant at 12,000 total steps.  

We have compared the performance of MLE and taMSD analyses
using synthetic particle trajectories  both {without} localization noise (Figs. S1-S2)
and {with} localization noise (Figs. S3-S4). 
The detailed procedures involved in the MLE analysis and the taMSD analysis are given
in Methods (Sec. \ref{methods}).
For confined diffusion (Figs. S1 and S3), MLE outperforms taMSD.
Even though both the MLE and the taMSD diffusivity estimates exhibit a positive
bias in their estimations of the diffusion coefficient,
both the bias and the error are significantly
less for MLE than for taMSD, especially in the presence
of localization noise.
When analyzing synthetic particle trajectories with localization noise,
 taMSD-based estimates of  the confinement length are  erratic.
By contrast, even with localization noise,  MLE yields reasonable confinement size estimates,
provided the reduced confinement size
($L_{reduced} = \frac{L}{\sqrt{12D\Delta t}}$) is sufficiently small.
As could be expected, the range of reduced confinement sizes for which MLE provides reasonable estimates
increases with increasing track length, because
the increased errors for larger reduced confinement sizes are associated with
each particle's limited sampling of its confinement, that is inevitable for short tracks.
 For the MLE analyses, 
 the static localization noise estimate
 was slightly negatively biased with a decreasing bias for increasing track length.

For fBm (Figs. S2 and S4) also, MLE is superior to taMSD.
In this case,
MLE and taMSD estimates both appear unbiased when particle tracks do not contain localization noise. However, the MLE estimates show noticeably lower errors.
In the presence of 
localization noise, both MLE and taMSD estimates for the diffusivity, anomalous exponent, and static localization noise become biased.
However, both the bias and the error are considerably less for MLE than for taMSD.
As expected, the bias and the error are reduced the longer the trajectories analyzed
both without and with localization noise.

These collected results unambiguously demonstrate that MLE improves upon taMSD
estimates for non-normal diffusion modes, in each case
reliably characterizing the underlying diffusion model over a broader range of parameter space.
They also emphasize that the presence of static localization noise reduces the quality of both taMSD-
and MLE-based estimation, and in some cases, may introduce a bias,
underscoring the importance of properly incorporating the effect of localization noise.
As expected, bias and errors are reduced for longer (but fewer) individual trajectories,
even for a fixed total number of time steps.

\subsection{\label{mleBIC}Performance of Bayesian model selection to classify individual particle trajectories}

\begin{figure*}[!htbp]
\centering{\includegraphics[width = 5in]{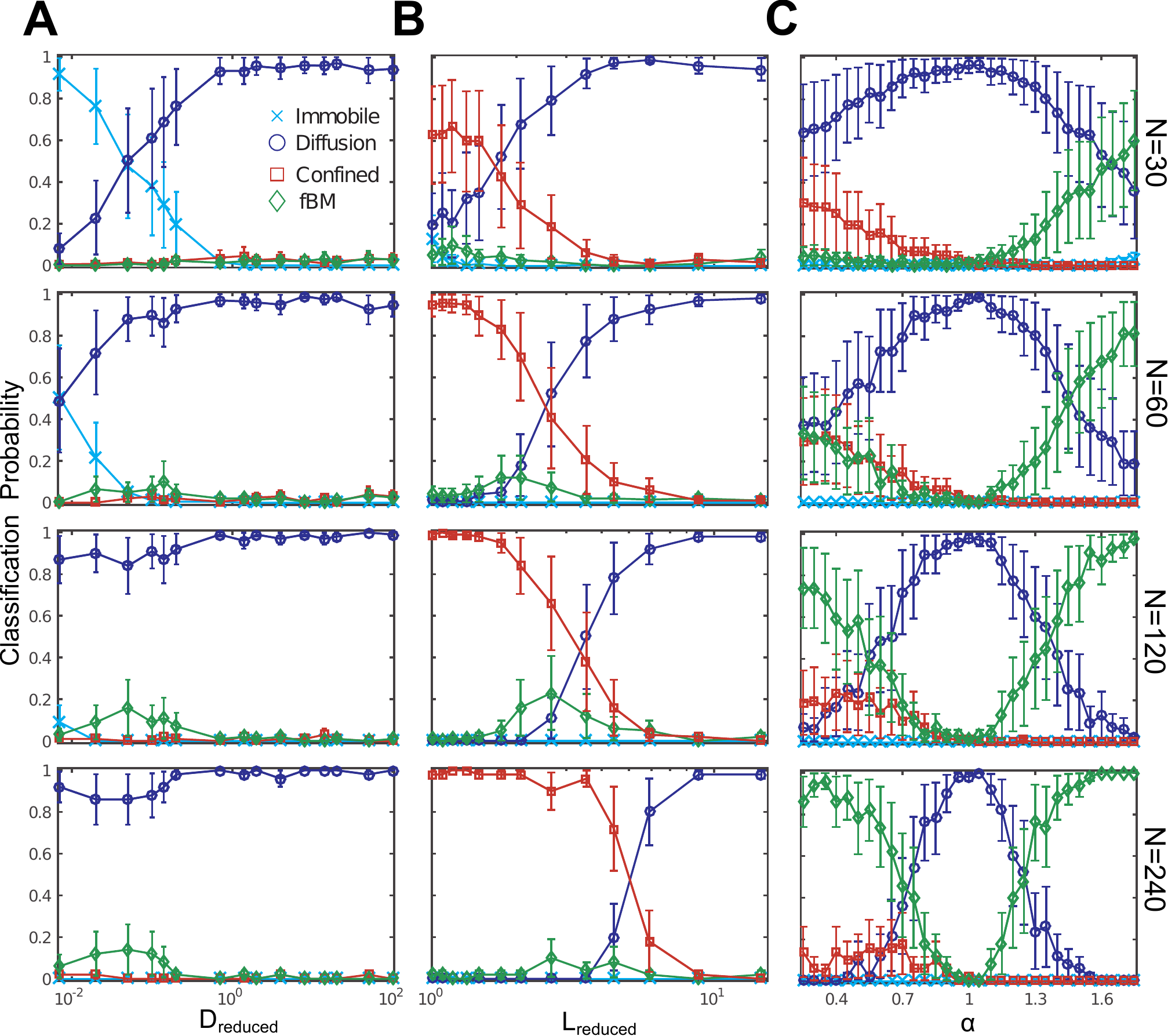}}
\caption{\label{fig:noiseBIC} \small {
Classification probability via mleBIC of simulated particle trajectories \emph{with} localization noise for various particle track lengths,
 undergoing (A) normal diffusion for various underlying diffusivities, (B) confined diffusion for various reduced confinement sizes, and (C) fractional Brownian motion for various anomalous exponents. 
 Each row represents a different particle track length: (first row) $N=30$ steps, (second row) $N=60$ steps, (third row) $N = 120$ steps, and (last row) $N = 240$ steps.}
 The probability of each model was calculated on the basis of the fraction of tracks classified to that model
  at each point in parameter space, and  is specified by a unique marker and color: immobile (cyan cross), normal diffusion (blue circles), confined diffusion (red square), and anomalous diffusion (green diamond).
   Error bars represent the observed standard deviation.  
}
\end{figure*}

For experimental particle trajectories, the underlying mode of diffusion is in general unknown \textit{a priori}.
Therefore, some criterion must be imposed to select the best model, {\em i.e.}
to statistically assess which diffusion model best describes any given particle trajectory.

According to Bayesian model selection, classification can be made by inferring the probability of the $k$th diffusion model, $\mathcal{M}_k$, from a  trajectory, $P(\mathcal{M}_k|\Delta {\bf x})$, where $\Delta {\bf x}$ represents the vector of displacements from a particle trajectory. According to Bayes' rule, the probability of diffusion model  $k$ is:  
$P(\mathcal{M}_k|\Delta {\bf x}) = \frac{P(\Delta {\bf x}|\mathcal{M}_k)P(\mathcal{M}_k)}{P(\Delta {\bf x})}$,
where $P(\mathcal{M}_k)$ is the model prior, $ P(\Delta {\bf x})$ may be viewed as a normalization constant,
given by $\sum_{i\in\mathcal{M}} P(\Delta {\bf x}|\mathcal{M}_i)P(\mathcal{M}_i)$, and $P(\Delta {\bf x}|\mathcal{M}_k)$ is the model evidence given by $P(\Delta {\bf x}|\mathcal{M}_k) = \int P(\Delta {\bf x}|{\bf \theta},\mathcal{M}_k)  P({\bf \theta}|\mathcal{M}_k) d{\bf \theta}$, where $P(\Delta {\bf x}|{\bf \theta},\mathcal{M}_k)$ is the likelihood distribution and $P({\bf \theta}|\mathcal{M}_k) $ is the prior distribution of the parameters, ${\bf \theta}$, of model $k$. 

Although the model prior, $P(\mathcal{M}_k)$, may be specified to express a preference for a particular model,
we elect to take an agnostic approach and assume that all diffusion models are equally probable.
In this manner, the model evidence is the only term of interest as the normalization 
absorbs all other contributions.  However, the priors of each model, $P({\bf \theta}|\mathcal{M}_k) $, may introduce a bias which becomes more pronounced when the peak of the likelihood distribution is not sharp. A representative likelihood distribution for confined diffusion  (Fig. S5), and fBm  (Fig. S6), calculated using simulated data for various track lengths, demonstrates that the likelihood distribution for these non-normal diffusion modes
is indeed broad near its global maximum. 

To minimize the influence from priors, we employ a Laplace approximation to the model evidence and assume a broad multivariate Gaussian prior with a full rank covariance matrix, which leads
via standard manipulations
to the Bayesian information criterion given according to  \cite{bishop2006pattern, murphy2012machine}:
\begin{equation}
\label{eqn:BIC}
\rm{BIC} = \ln P(\Delta {\bf x}|\mathcal{M}_k) = \ln P(\Delta {\bf x}|\hat{\mathbf{\theta}},\mathcal{M}_k) - \frac{N_{params}}{2} \ln M,
\end{equation}
 where  $\hat{ \theta}$ are the maximum likelihood parameters of model k, $N_{params}$ is the number of free parameters, and $M$ is the number of particle track displacements.  
  
In summary, for a given trajectory, the MLE is found for each candidate diffusion model, according to Methods (Sec. \ref{methods}),  yielding the parameter estimates and log-likelihood value, from which the BIC can be calculated
(Eq. \ref{eqn:BIC}).
The model probability for each diffusion model can be subsequently calculated according to:
\begin{equation}
\label{eqn:BICprob}
P(\mathcal{M}_k| \Delta {\bf x}) = \frac{ \exp{ \left( \rm{BIC}_k - \overline{\rm{BIC}}\right) } }{\sum_{i=1}^K \exp{\left(\rm{BIC}_k - \overline{\rm{BIC}} \right)}} ~,  
 \end{equation}
where $\overline{\rm{BIC}}$ is the maximum BIC value across $K$ candidate diffusion models.  Thus, classification is determined by the diffusion model which yields the highest model probability. Henceforth, this analysis pipeline is referred to as mleBIC.

To understand the statistical limits of classification under ideal circumstance, namely particles which have constant diffusion properties throughout the duration of their trajectories, we employed mleBIC across various sets of synthetic particle trajectories \emph{with} static and dynamic localization noise for each canonical diffusion mode (Fig. \ref{fig:noiseBIC}).  
For short particle trajectories undergoing normal diffusion (Fig. \ref{fig:noiseBIC}A), a normal diffusion model was favored with a high probability when $D_{reduced} = \frac{D\Delta t}{\sigma^2} > 1$.  When $D_{reduced} < 1$, the underlying static localization noise dominates the underlying diffusion, which leads mleBIC to favor an immobile model. Thus, more statistics are necessary  to reject the simpler immobile model.  

For particle trajectories undergoing confined diffusion (Figs. \ref{fig:noiseBIC}B), when confinement sizes are small, a confined diffusion model is favored.  As the confinement size increases, a normal diffusion model becomes favored.  At this confined-to-normal crossover, a small preference for anomalous diffusion is found.  As expected, longer trajectories provide more opportunities to explore the boundaries of confinement, resulting in a wider region of parameter space for which a confined diffusion model is favored.  

 For particle trajectories undergoing fBm (Fig. \ref{fig:noiseBIC}C), a normal diffusion model is mostly favored when particle trajectories are short ($N \leq 60$). A fBm diffusion model is not consistently favored until trajectories contain 240 steps, albeit only when the anomalous exponent is below 0.7 or greater than 1.3.  
As expected, when particle trajectories contain minimal localization noise errors, mleBIC yields improved estimation for fBM (Fig. S7). Thus, the presence of localization noise requires even longer tracks for proper classification, even though MLE can determine reliable estimates for the underlying diffusivity and anomalous exponent (Fig. S4).
BIC's built-in parsimony causes it to favor a normal diffusion model,
when there is not enough data to support a non-normal diffusion model,
even when the correct model corresponds to non-normal diffusion.
This behavior seems not undesirable.

Similar to taMSD analysis, mleBIC does not take into account transitions between diffusive states.  While analyzing subsets of the data may allow for different diffusive states within a particle trajectory, figures \ref{fig:noiseBIC} and S7 illustrate that accurate classification cannot be made for wide ranges of parameter space, even in the most ideal circumstances. As the trajectories become longer, the statistical power grows, thereby allowing for improved mleBIC classification over a wider parameter space, and misclassification gradually reduces. However, longer particle trajectories which have constant diffusion properties becomes increasingly unlikely, especially when a particle is diffusing in a complex environment such as a living cell. Thus, while mleBIC is certainly an improvement over taMSD analysis, the statistical power of classification by analyzing particle trajectories on an individual basis remains limited.


\subsection{\label{pEM}Systems-level analysis of a collection of particle trajectories}

To augment the limited statistics provided by individual particle trajectories,
pEM simultaneously analyzes a collection of trajectories by employing a systems-level likelihood function 
to account for a finite number of unique diffusive states,
each of which we envision to arise as a result of particular interactions within the cell.
Here, we extend the original pEM framework \cite{Koo} to now include non-normal modes of diffusion,
{\em i.e.} we lift the short-time-diffusion approximation.
A powerful aspect of this new version of pEM  is that it is essentially a {\em  model-free approach},
in that no prior assumptions need be made, concerning which types of  diffusion mode are present in the data
at hand.
 
To implement the new version of pEM,  we first write the systems-level  log-likelihood function:
\begin{equation}
\label{eqn:lnLnoise}
\ln{\mathcal{L}(\Delta \hat{\mathbf{x}}|\hat{\boldsymbol{\pi}},\hat{\mathbf{\Sigma}})}  = \sum_{m=1}^{M} \ln{\left\{ \sum_{k=1}^K\pi_k P(\Delta \mathbf{x}_m|\mathbf{\Sigma}_k) \right\}},
\end{equation}
where $M$ is the total number of tracks, which collectively realize $K$ distinct underlying diffusive states,
$\Delta \mathbf{x}_m$ represents the vector of $N_m$ displacements for particle trajectory $m$, $\Delta \mathbf{x}_m = \{\Delta x_m(n)\}_{n=1}^{N_m}$, $\Delta \hat{\mathbf{x}} = \{\Delta \mathbf{x}_m\}_{m=1}^M$ is the set of $M$ particle track displacements, $\hat{\boldsymbol{\pi}} = \{ \pi_k\}_{k=1}^K$ is the set of variables which represent the fraction of the population of trajectories that realize diffusive state $k$, which is bounded and normalized:  $0 \leq \pi_k \leq 1$ and $\sum_{k=1}^K \pi_k = 1$, and $\hat{\mathbf{\Sigma}} = \{\mathbf{\Sigma}_k\}_{k=1}^K$ is the set of covariance matrices which defines each diffusive state.

Importantly, the theoretical covariance matrix for any diffusion mode that undergoes a stationary Gaussian process, including in the presence of localization noise,
has a symmetric Toeplitz form (Table \ref{table:covariancenoise}), so that element $(i,j)$ of the covariance matrix depends only on $|i-j|$.
We can impose the requirement that the covariance matrix for each diffusive state, $\mathbf{\Sigma}_k$,
take on such a symmetric Toeplitz form by averaging the diagonal, one-off-diagonal, two-off-diagonal, {\em etc.}
 elements of the empirical covariance matrix for particle track $m$ to obtain the experimental covariance matrix elements for track $m$:
$\mathbf{C}_m(i,j)=\mathbf{C}_m(|i-j|) = \langle \Delta \mathbf{x}_m(l) \Delta \mathbf{x}_m(l+|i-j|) \rangle$,
where the average is taken over all possible values of $l$ for track $m$. 

Furthermore, because the covariance structure of each diffusion mode decreases rapidly to zero
for increasing separation between displacements --  {\em i.e.} with increasing $|i-j|$ --
 we can reasonably restrict the number of informative covariance matrix elements
 that we include in the analysis
 by setting $\mathbf{C}_m(\vert i-j \vert) = 0$, for $|i-j| > f$, where $f$ is the number of off-diagonal covariance
 matrix elements included in the analysis.
 If $f=0$, only the diagonal elements of the covariance matrix are
 permitted to be non-zero, reproducing the theoretical structure of the covariance matrix for simple diffusion
 in the absence of localization noise. 
 For $f=1$, one-off-diagonal element is included, permitting the covariance matrix to properly account for localization noise sources.
In principle, different diffusive states, which are characterized by unique diffusion properties, may be distinguished one from another on the basis of different values of the covariance matrix elements.
The inclusion of additional off-diagonal terms introduces additional information to help distinguish diffusive states that undergo confined diffusion, fBm or other modes of non-normal diffusion.

Because pEM discovers the values of these covariance matrix elements for each diffusive state,
it is not necessary to specify ahead of time what diffusion modes are present, beyond specifying
$f$. It is in this sense that this version of pEM is model independent.
In the case of  $K$ diffusive states, insisting that the covariance matrix must be
a symmetric Toeplitz matrix and limiting the number of
off-diagonal matrix elements to $f$ means that the number of model parameters is equal to  $K(1 + f)+K-1$.
(There are $K-1$ independent population fractions.)

Maximizing Eq. \ref{eqn:lnLnoise} with respect to $\{\mathbf{\Sigma}_k, \pi_k\}_{k=1}^K$
naturally yields the expectation-maximization (EM) algorithm \cite{dempster1977maximum}.
 In the expectation step, the posterior probability,
 $\gamma_{mk}$, that particle trajectory $m$ realizes diffusive state $k$, given
  the current estimates for  $\mathbf{\Sigma}_k$,  and $\pi_k$, 
 is calculated according to:
\begin{equation}
\label{eqn:gamma}
\gamma_{mk} =\frac{\pi_kP(\Delta \mathbf{x}_m|\mathbf{\Sigma}_k) } { \sum_{j=1}^K\pi_j P(\Delta \mathbf{x}_m|\mathbf{\Sigma}_j)}.
\end{equation}
In the maximization step, the posterior probability is used to update the parameter estimates of each diffusive state:
\begin{eqnarray}
\label{eqn:Dupdate}
\mathbf{\Sigma}_k &=& \frac{1}{M_k} \sum_{m=1}^M  \gamma_{mk} \mathbf{C}_m \\
\pi_k &=&  \frac{M_k}{M}
\end{eqnarray}
where $\mathbf{C}_m(i,j) = \langle \Delta \mathbf{x}_m(i) \Delta \mathbf{x}_m(j) \rangle$ and 
$M_k = \sum_{m=1}^{M} \gamma_{mk}$. 
The EM algorithm solves these equations iteratively until the change in the log-likelihood
becomes smaller than a set threshold \cite{dempster1977maximum}.

The extension to higher dimensions than one is carried out  as follows.
We calculate the expectation step by averaging the posterior probability over each dimension using the same parameter estimates.  For the maximization step, the maximized parameter estimates are calculated separately for
 each dimension and then averaged.  At each step in the iteration procedure, the complete log-likelihood is calculated by summing the log-likelihood from each dimension.
 
Although the EM algorithm guarantees convergence to a maximum \cite{dempster1977maximum},
convergence to the global maximum is not guaranteed, depending on the initial parameter values.
However, 
as described in detail in Ref.~\cite{Koo} and summarized in the Methods (Sec. \ref{methods}),
suitably perturbing the likelihood surface,  namely pEM, is a
computationally efficient means to reach the global maximum likelihood.

Since the number of diffusive states is not known {\it a priori},
we repeat the pEM procedure for different numbers of diffusive states,
finding the maximum likelihood in each case.
To maintain model parsimony,
we again employ the Bayesian Information Criterion to penalize for the inclusion of additional diffusive states,
via a systems level extension of Eq. \ref{eqn:BIC}.
Specifically, we select the model with the largest value of the systems-level BIC,
where now $\log \mathcal{L}$ is the systems-level likelihood function (Eq. \ref{eqn:lnLnoise}),
the number of free parameters is $N_{params} = K(1+f)+K-1$,
and $M$ is the total number of particle track displacements across the population of tracks.

\begin{figure*}[t!]
\centering{\includegraphics[width = 6.5in]{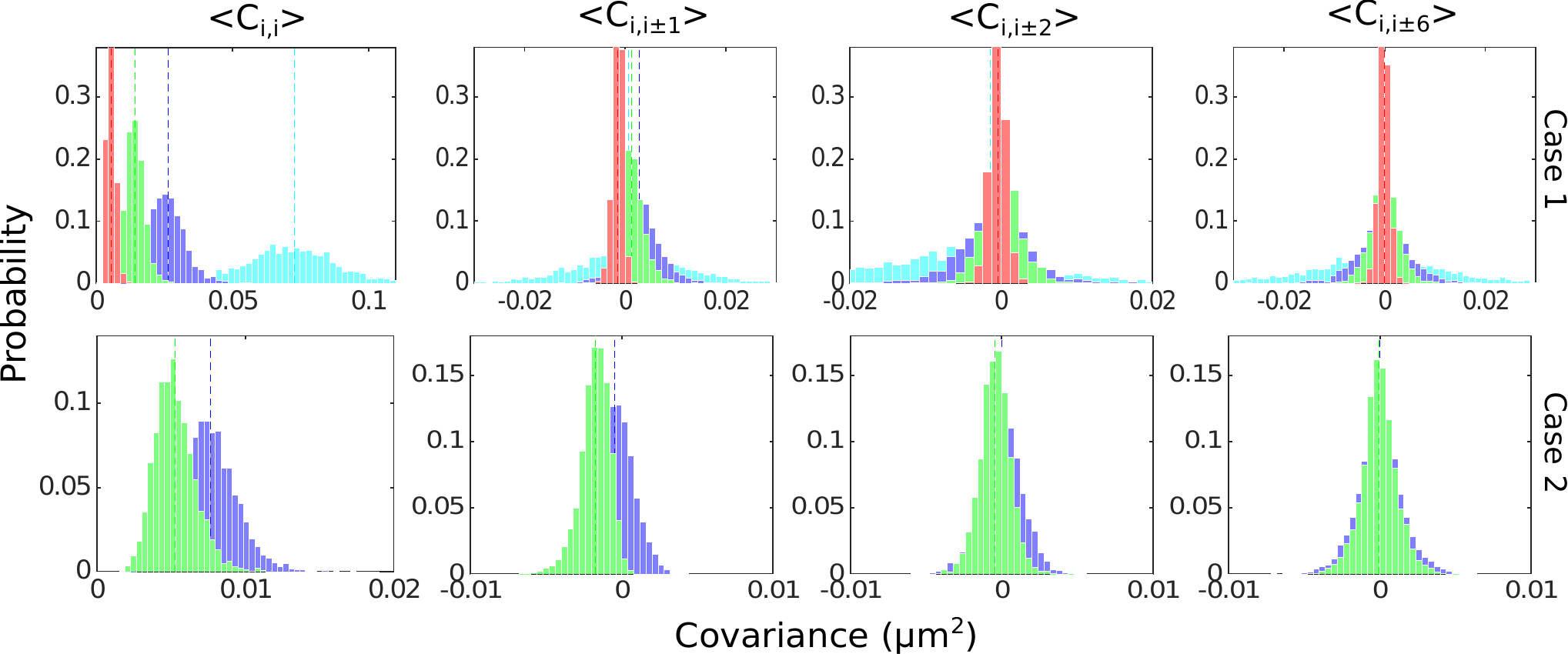}}
\caption{\label{Case1_1} \small {
Empirical probability distributions of the mean covariance matrix elements, $\langle {\bf C}(i,j) \rangle$, for $j=i$,  $j = i\pm1$, 
$j=i\pm2$, and  $j=i\pm6$ for case 1 (top row), where  states  1, 2, 3, and 4 are shown in red, green, blue, and cyan,
respectively, and for case 2 (bottom row), where states 1 and 2 are shown in green and blue, respectively.
Vertical dashed lines indicate the theoretical values with a color corresponding to each diffusive state.
}
}
\end{figure*}

The procedure described so-far makes the assumption that the
diffusive properties remain constant throughout the duration of each trajectory.
In order to extend pEM, so that it can be applied to trajectories containing transitions
 between different diffusive states, we split each trajectory into equal-size bins,
 such that each bin contains $B$ sequential steps.
The assumption of a constant covariance matrix is still assumed to hold within each such bin,
but different bins can realize different diffusive states.
In this way, pEMv2 is able to account for
transitions between different diffusive states within the overall  trajectory.
Each bin is treated as a Markovian measurement of the diffusive state, Eq. \ref{eqn:likelihood}.
The temporal resolution corresponds to the bin size.
 
To summarize, our enhanced version of pEM, which we call pEMv2,
examines a population of binned particle trajectories, each containing $B$ steps,
to determine the number of unique covariance matrices, contained in the population.
It accomplishes this goal
by classifying each binned trajectory to a particular diffusive state, based on similarities in the covariance structure
among trajectories.
Using the resultant classification, pEMv2  then updates the parameter estimates for each diffusive state.
Iteration of this process allows pEMv2 to learn  in an unsupervised manner
what  unique covariance structures, \textit{i.e.} what diffusive states,
are realized within the population of binned trajectories.
 Since the number of diffusive states is intrinsically handled by the BIC (Eq. \ref{eqn:BIC}),
 the user-controllable  parameters for pEMv2,
are the number of off-diagonal elements to include in the covariance matrix, $f$, and the bin size, $B$.

\begin{table}[b!]
\small
\centering
\begin{tabular}{|c|c|cccc|}
\hline
\multicolumn{1}{|m{.5cm}}{} & \multicolumn{1}{c|}{} & \multicolumn{1}{m{1.25cm}}{~~~~1} & \multicolumn{1}{m{1.25cm}}{~~~~2} & \multicolumn{1}{m{1.25cm}}{~~~~3} & \multicolumn{1}{m{1.25cm}|}{~~~~4}\\
\hline
 \parbox[t]{2mm}{\multirow{6}{*}{\rotatebox[origin=c]{90}{Case 1}}}
& mode & Confined & Normal & fBM & fBM \\
&$D_k^{sim}$ ($\mu\mathrm{m}^2\mathrm{s}^{-1}$) & 0.05 & 0.15 & 0.25 & 0.4\\
&$L_k^{sim}$ ($\mu\mathrm{m}$) & 0.13 & & & \\
&$\alpha_k^{sim}$ & 1 & 1 & 0.9 & 0.6 \\ 
&$\sigma_k^{sim}$ ($\mu\mathrm{m}$)&0.04&0.04&0.04&0.04\\
&$\pi_k^{sim}$&0.25&0.25&0.25&0.25\\
\hline
 \parbox[t]{2mm}{\multirow{6}{*}{\rotatebox[origin=c]{90}{Case 2}}}
& mode & Confined & Normal & &\\
&$D_k^{sim}$ ($\mu\mathrm{m}^2\mathrm{s}^{-1}$) & 0.06 & 0.06 &&\\
&$L_k^{sim}$ ($\mu\mathrm{m}$) & 0.1 & && \\
&$\alpha_k^{sim}$ & 1 & 1 &&\\ 
&$\sigma_k^{sim}$ ($\mu\mathrm{m}$)&0.04&0.04 &&\\
&$\pi_k^{sim}$&0.4&0.6 &&\\
\hline
 \parbox[t]{2mm}{\multirow{6}{*}{\rotatebox[origin=c]{90}{Case 3}}}
& mode & Confined & Confined & Normal & \\
&$D_k^{sim}$ ($\mu\mathrm{m}^2\mathrm{s}^{-1}$) & 0.005 & 0.1 & 0.3 &\\
&$L_k^{sim}$ ($\mu\mathrm{m}$) & 0.05 & 0.2 & &\\
&$\alpha_k^{sim}$ & 1 & 1 & 1  &\\ 
&$\sigma_k^{sim}$ ($\mu\mathrm{m}$)&0.04&0.04&0.04 &\\
&$\pi_k^{sim}$&0.33&0.33&0.34 &\\
\hline
 \parbox[t]{2mm}{\multirow{6}{*}{\rotatebox[origin=c]{90}{Case 4}}}
 & mode & Normal & fBM & Normal & fBM \\
& $D_k^{sim}$ ($\mu\mathrm{m}^2\mathrm{s}^{-1}$) & 0.001 & 0.03 & 0.2 & 0.45\\
& $L_k^{sim}$ ($\mu\mathrm{m}$) & & & & \\
& $\alpha_k^{sim}$ & 1 & .7 & 1 & 0.9 \\ 
& $\sigma_k^{sim}$ ($\mu\mathrm{m}$)&0.04&0.04&0.04&0.04\\
& $\pi_k^{sim}$&0.25&0.25&0.25&0.25\\
\hline
\end{tabular}
\caption{Simulation parameters for synthetic particle trajectories generated for case 1, case 2,
case 3, and case 4.
}
\label{Table2}
\end{table}

\subsubsection{\label{numFeatures}Dependence on the number of covariance terms}

 To investigate the performance of pEMv2,
 we have generated a number of sets of synthetic particle trajectories
 containing different numbers of diffusive states
 and different degrees of similarity between the covariance terms across diffusive states.
 Table \ref{Table2} specifies the four sets of diffusion parameters (case 1 through case 4),
 which were used to generate the synthetic data sets.
 There are no transitions among different diffusive states for case 1 and case 2, {\it i.e.} the transition probability matrix (${\bf A}$) is given by ${\bf A} = \delta_{i,j}$.
However, for case 3 and case 4, transitions are permitted
with
the corresponding matrices of transition probabilities given by
\begin{eqnarray}
{\bf A}_3 = \left( \begin{array}{ccc}
0.995 & 0.001 & 0.004 \\
0.001 & 0.995 & 0.004 \\
0.015 & 0.015 & 0.970 \end{array} \right)
\label{AAA333}
\end{eqnarray}
for case 3, and
\begin{eqnarray}
{\bf A}_4 = \left( \begin{array}{cccc}
1-3p & p & p & p  \\
p & 1-3p & p & p \\
p & p & 1-3p& p \\
p & p & p & 1-3p \\
 \end{array} \right)
 \label{AAA444}
\end{eqnarray}
for case 4,
where  $p$ is input into the simulation selected from one of $\{0,~0.003, ~0.005,~0.01,~0.015,~ 0.02, ~0.03 \}$.

The  covariance matrix elements of different diffusive states must be sufficiently distinct in order for pEMv2 to resolve
them as separate diffusive states.
First, therefore,
we sought to explore the effect of the number of off-diagonal covariance matrix elements ($f$), 
that are included in pEMv2 analysis.
Figure \ref{Case1_1} shows the measured probability distributions of the average covariance matrix elements, $\langle {\bf C}(i,j) \rangle$
for $|i-j|= 0$, 1, 2, and 6, determined from populations containing
1,500 synthetic trajectories, realizing four diffusive states
 with diffusion parameters corresponding to case 1 (top row),
and
1,500 synthetic trajectories, realizing two diffusive states with diffusion parameters corresponding to case 2
(bottom row).
To recapitulate the variability found experimentally,
the trajectory lengths were distributed according to an exponential probability distribution
with a characteristic length of 25 steps,
with a minimum cut-off of 15 steps and a maximum cut-off of 60 steps.
In addition, because there are no transitions,  in our analyses of case 1 and case 2,
 we analyzed each complete trajectory as a whole, as in the original version of pEM,
 without splitting into bins.
 Case 1 corresponds to four diffusive states, 
 two normal diffusion, one fBM, and one confined diffusion but their diffusion coefficients are well-separated from each other.  
Case 2 corresponds to two diffusive states with the same
 diffusion coefficient, one corresponding to normal diffusion and the other to confined diffusion.

For both case 1 and case 2, the means of the distributions of $\left < {\bf C}(i,i) \right >$ and $\left < {\bf C}(i,i\pm1) \right >$
for each diffusive state are well separated.
For case 1, however, the means of $\left < {\bf C}(i,i\pm2) \right >$ are all very similar to each other and are close to zero, with the exception of state 4 (cyan).  By contrast, for case 2, the means of $\left < {\bf C}(i,i\pm2) \right >$ for state 1 and state 2 remain distinguished from each other. For case 1 and case 2, the means of $\left < {\bf C}(i,i\pm6) \right >$
 for each diffusion state are all very similar to each other and are all close to zero, albeit their widths remain distinct.

\begin{figure}[t!]
\centering{\includegraphics[width = 2.5in]{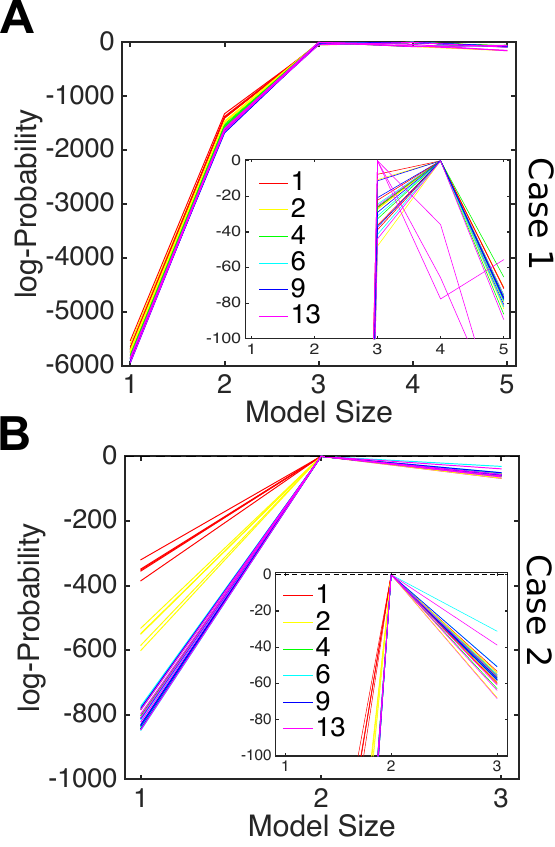}}
\caption{\label{Case1_2} \small {
Log-probability as a function of the number of diffusive
states (model size) for (A) case 1 and (B) case 2,
determined by pEMv2 analysis using $f=1, 2, 4, 6, 9,$ and 13 off-diagonal covariance matrix elements,
shown in red, yellow, green cyan, blue, and magenta, respectively. Each data set consists of 5 sets of simulated particle tracks (shown as a different curve) for each $f$ (shown as a different color)}. The inset shows a zoomed in representation near the maximum log-probability. 
}
\end{figure}

Fig. \ref{Case1_2} shows the log-probability of each model size,
determined
on the basis of
BIC score (Eq. \ref{eqn:BICprob}),
as a function of model size for different numbers of non-zero
off-diagonal covariance matrix elements between $f=1$ and 13.
For case 2, the correct number of diffusive states is found ($K=2$), irrespective of $f$.
For case 1, where $K=4$, pEMv2 is able to successfully determine
the correct numbers of diffusive states,  as indicated by the maximum log-probability,
except when $f=13$, for which a 3 diffusive state model is favored for three out of the five data sets analyzed.

\begin{figure}[t!]
\centering{
\includegraphics[width = 3.5in]{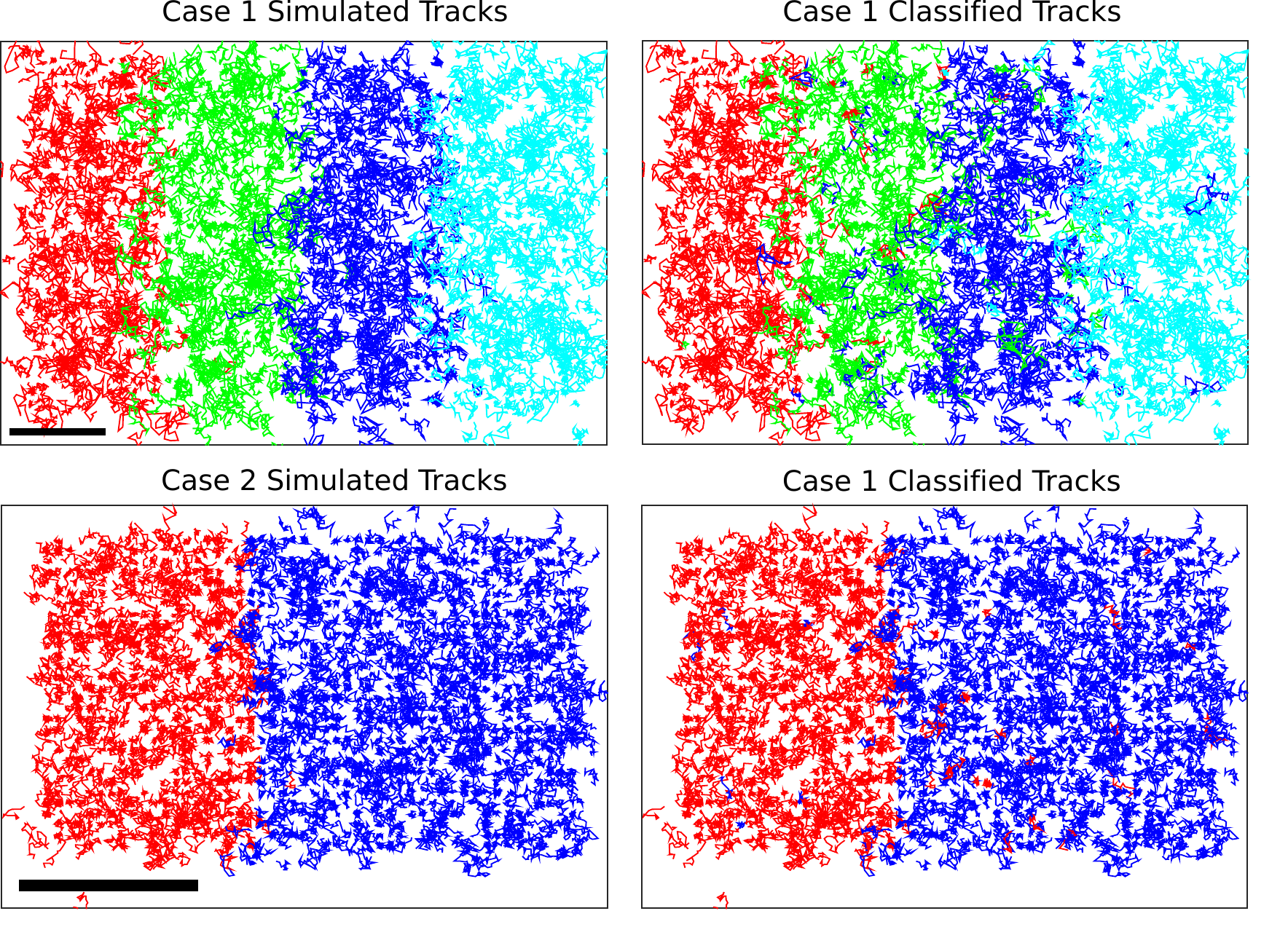}
}
\caption{\label{Case1_4} \small {
Representations of 1500 synthetic particle trajectories for case 1 (top row) and case 2 (bottom row).
In the left column, each trajectory is depicted using a color, corresponding to the the
known, simulated diffusive state of the track.
In the right column,  each trajectory is depict using a color, corresponding to the diffusive state, that yields
maximum posterior probability, determined on the basis of pEMv2 using $f=6$ off-diagonal covariance
matrix elements.
The starting position of each trajectory is sequentially placed on a 2-dimensional grid, separated one from another
by $1 ~\mu\rm{m}$
(top row) and $0.4~\mu\rm{m}$ (bottom row). In both cases, the scale bar represents 5 $\mu$m.
}
}
\end{figure}

A visual representation of how successfully pEMv2 determines the correct diffusive state is
given in
Fig.~\ref{Case1_4}, which shows
1500 synthetic particle trajectories corresponding to  case 1 (top row) and case 2 (bottom row).
In the left column, each trajectory is depicted using a color, corresponding to the 
known, simulated diffusive state of the track.
In the right column,  each trajectory is depicted using a color, corresponding to the diffusive state that realizes
the maximum posterior probability for that track, determined using $f=6$ off-diagonal covariance
matrix elements for case 1 and case 2.
Although there are a few misclassified trajectories,
the overwhelming majority of the trajectories are correctly classified,
demonstrating that pEMv2 is capable of
reliably uncovering the diffusive states in these cases.

\begin{figure}[t!]
\centering{\includegraphics[width = 2.5in]{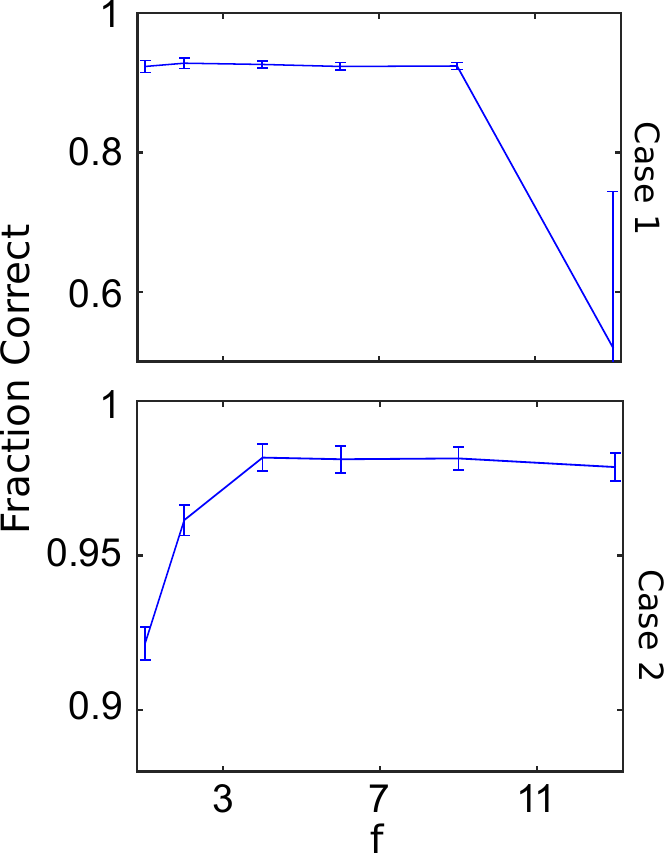}}
\caption{\label{Case1_3} \small {
Fraction of trajectories classified into the correct diffusive state as a function of the number of off-diagonal
covariance matrix elements used in the pEMv2 analysis of case 1 (top) and case 2 (bottom).
Error bars represent the observed standard deviation across 5 different sets of simulated particle tracks.
}
}
\end{figure}

Fig.~\ref{Case1_3} shows the fraction of correctly classified trajectories
as a function of the number of off-diagonal covariance matrix elements,
confirming that pEMv2 reliably classifies trajectories to the correct diffusive state.
The classification accuracy
shows only a modest dependence on the number of off-diagonal covariance matrix elements included in the analysis:
For case 1, the accuracy of classification is uniformly high for $f$ between 1 and 9,
suggesting that the first off-diagonal covariance matrix element ($f=1$) is decisive
in case 1.
The decrease in classification accuracy for $f=13$ may be because of the inclusion in this case of a
large number of noisy off-diagonal matrix elements, suggesting that it is preferable to not include too many off-diagonal
covariance matrix elements.
For case 2, the accuracy noticeably improves as $f$ increases from 1 to 4, and remains high thereafter,
suggesting that off-diagonal covariance matrix elements up to $f=4$ are informative for classification in this case.

\begin{figure*}[t!]
\centering{\includegraphics[width = 6.5in]{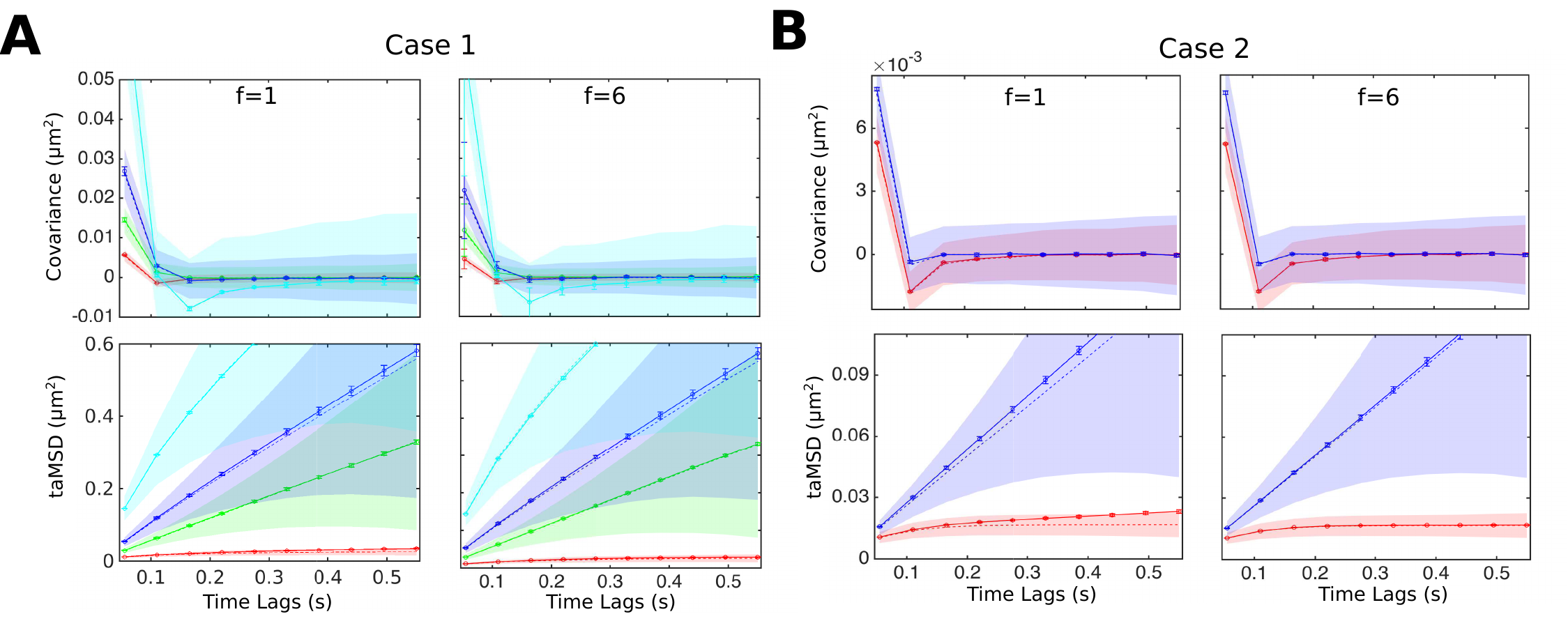}}
\caption{ \label{Case1_5}
Average covariance matrix elements and average taMSD from maximum posterior classification as a function of time lag for (A) states 1, 2, 3, and 4 of case 1,  represented in red, green, blue and cyan, respectively, and (B) states 1 and 2 of case 2, represented in red and blue, respectively.
 Each data point represents the average over five different sets of simulated particle tracks, analyzed using pEMv2
 using $f=1$ (left column) and $f=6$ (right column). The solid lines linking the data points are guides-to-the-eye.
 The error bars correspond to the standard deviation of the mean across 5 trials.
Shown as the dashed curves are the true matrix elements and the true ensemble-averaged taMSD for each state, determined using the known diffusive states of trajectories, while the shaded bands represent its standard deviation.
}
\end{figure*}

The classified covariance matrix elements and the
 classified taMSD are shown in Fig.~\ref{Case1_5} for each diffusive state corresponding to case 1.
In this instance, using either $f=1$ or $f=6$ in the analysis leads to the characterization of
each diffusive state with high fidelity,
with the measured covariance matrix elements and measured taMSDs for each diffusive state, shown as the data points and the solid lines
in the figure,
almost exactly matching the corresponding true covariance matrix elements and true taMSDs,
shown as the dashed lines, which are very nearly coincident with the solid lines.
In comparison with mleBIC, which was unable to reliably classify 60-step trajectories undergoing fBm
with anomalous exponents of either $\alpha=0.9$ or even $\alpha = 0.6$, 
it is striking that pEMv2 is not only able to identify these two diffusive states (states 3 and 4 of case 1)
and to accurately categorize individual trajectories into these
states (Fig.~\ref{Case1_4}), pEMv2 is also able to accurately capture the anomalous behavior of their taMSDs (Fig.~\ref{Case1_5}).
Thus, the systems-level strategy employed by pEMv2 can find subtle deviations from non-normal diffusive behavior,
that are statistically challenging to uncover,
if trajectories are analyzed on an individual basis.

These observations show that the particular value of $f$ used is not critical.
In practice, we suggest that
a reasonable way to pick $f$ is on the basis of the average covariance matrix elements themselves
(see Fig. \ref{Case1_5}): we suggest picking $f$ to correspond to the off-diagonal term of the ensemble-averaged covariance matrix elements that has essentially converged to zero.
This choice should ensure that
all informative covariance matrix elements are included in the analysis, but that unnecessary noise is excluded.
We ascribe the failure to select the correct model for $f=13$ to be the result of
including unnecessary noise.
For the simulations in this paper, $f=6$ is a reasonable choice.

\subsubsection{\label{transitions}Uncovering transitions by splitting tracks}

A population of experimental trajectories, that realizes multiple diffusive states,
is likely to contain at least a subset of trajectories, which contain transitions among the diffusive states.
The prevalence of transitions depends on their underlying kinetics, {\em i.e.} on the transition rates.
Our concept for extending our methodology to permit analysis of trajectories with transitions is to split these trajectories into shorter pieces.
If the duration of the resultant short trajectories is less than the typical lifetimes of relevant diffusive states, then each short trajectory will 
with high probability realize a single diffusive state throughout, and the methods described above remain applicable to determine the diffusive states within
the population of these short trajectories.

To investigate the feasibility of this concept,
we simulated particle tracks with three diffusive states, corresponding to case 3 in Table~\ref{Table2},
that transition among each other with transition rate matrix ${\bf A}_3$ (Eq.~\ref{AAA333}).
The protocol used for generating transitions is described in the Methods (Sec. \ref{methods}).
The corresponding lifetimes of states 1, 2, and 3 are 200 ($\sim$6.4 s), 200 ($\sim$6.4 s), and 33 ($\sim$1 s) steps, respectively. 
We then divided the simulated trajectories into sets of short trajectories containing 5, 10, 15, 20, 25, 30, 60, 90  or 120 steps,
respectively,
while keeping the total number of steps and hence the total  positional information constant at 12000 total steps across all trajectories.
We then applied the pEMv2 methods described above
 to each population of different-length short trajectories, implicitly assuming that each short trajectory remains in the
same diffusive state throughout.
The number of off-diagonal matrix elements used in the analysis was fixed at $f=6$.

\begin{figure}[!htbp]
\centering{\includegraphics[width = 3.4in]{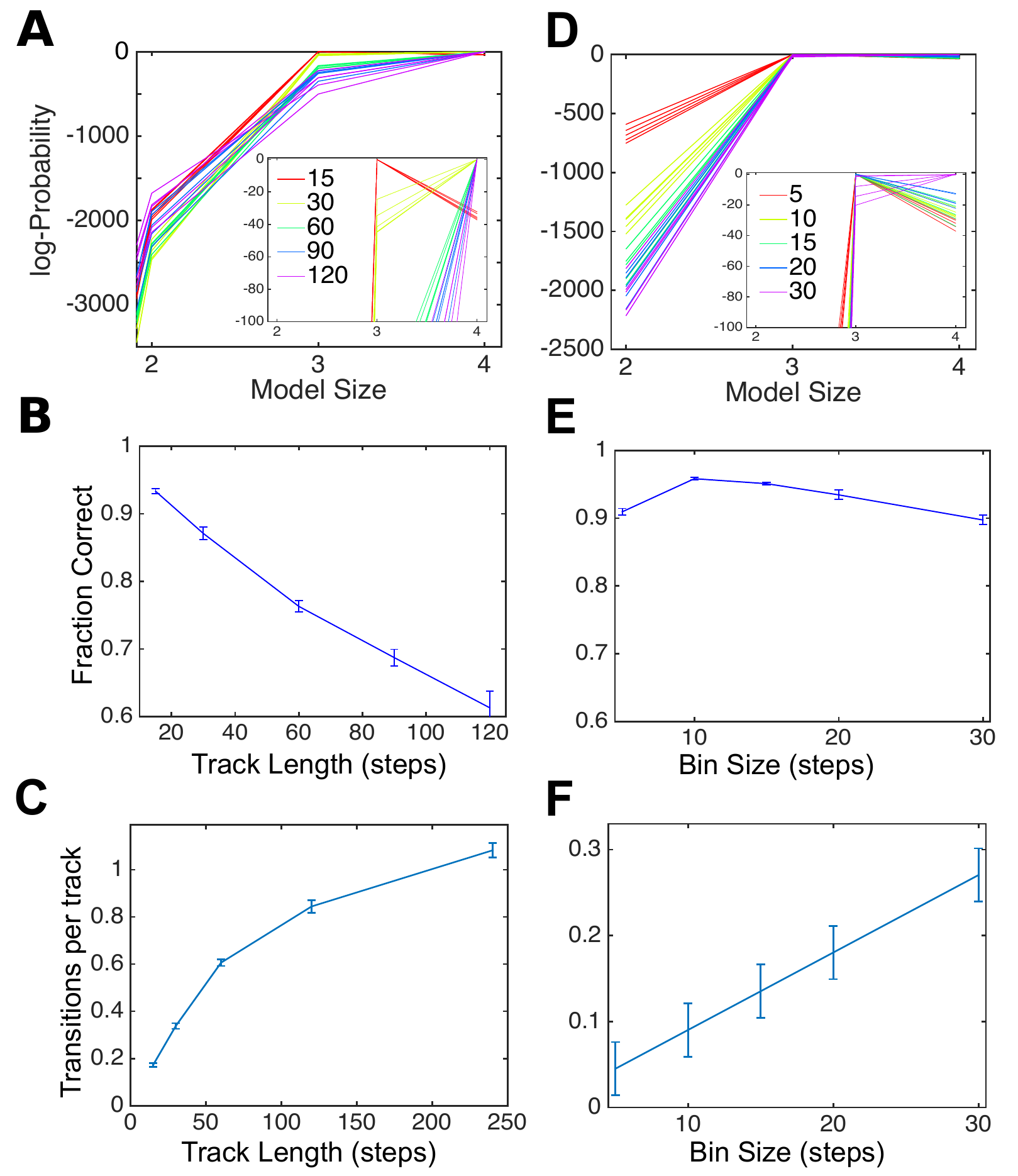}}
\caption{\label{Case2_1} \small {
Performance of pEMv2 on particle tracks that transition between diffusive states given by case 3. 
Log-probability determined by pEMv2 analysis for 5 sets of simulated particle tracks (shown as a different curve) with diffusive states given according to case 3 with  (A) track lengths of 15, 30, 60, 90, and 120 steps and (D) track lengths of 5, 10, 15, 20, and 30 steps created by splitting the 120 step data set in (A). Each simulation set is shown in a different color. 
Average fraction that the classified diffusive states matches the simulated diffusive state for (B) various track lengths and (E) various bin sizes.  
For the purposes of this comparison, when a bin contains a transition,
the ``true'' diffusive state is chosen to be the state with the highest number of displacements. 
Average transition rate per track for (C) various track lengths and (F) various bin sizes.
 (B,C,E,F) Error bars represent the observed standard deviation across the 5 data sets. 
}
}
\end{figure}

Fig.~\ref{Case2_1}A shows the BIC-based log-probability of various model sizes for simulated tracks with lengths  15, 30, 60, 90, and 120 steps. 
The log-probability selects the correct number of diffusive states ($K=3$) for only when $N =15$ steps.
For trajectories containing 30 or more steps, the BIC-based probability incorrectly favors a four diffusive state model, presumably in an effort to describe trajectories containing  transitions. 
Given that the three state model is correct, Fig.~\ref{Case2_1}B shows the fraction of the total number of steps that are assigned to the correct
diffusive state for each set of short trajectories, plotted as a function of the track length of each set.
Evidently,  the fraction of steps correctly assigned decreases as the trajectories became longer. 
 This trend is surely due to the fact that longer trajectories provide more opportunities to transition,
 as indicated by the increasing number of transitions per track with increasing trajectory length, shown in Fig.~\ref{Case2_1}C.  

\begin{figure}[!htbp]
\centering{\includegraphics[width = 3.5in]{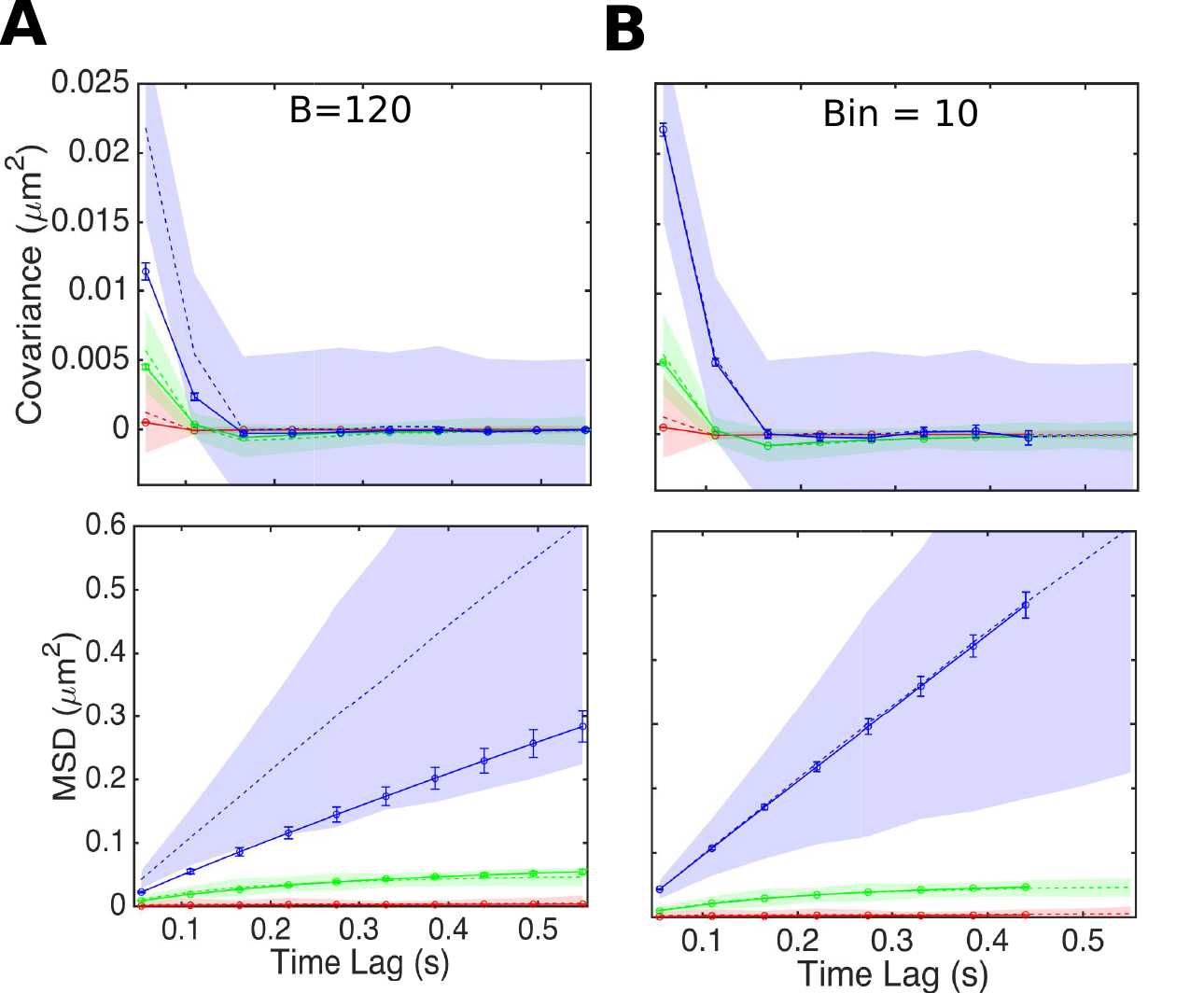}}
\caption{\label{Case2_2} \small {
Average covariance matrix elements and average taMSD from maximum posterior classification as a function of time lag for states 1, 2, and 3, represented in red green and blue, respectively, with a bin size of (A) 120 steps and (B) 10 steps.  
 Each data point represents the average over five different sets of simulated particle tracks, analyzed using pEMv2
 using $f=6$. The solid lines linking the data points are guides-to-the-eye.
 The error bars correspond to the standard deviation of the mean across 5 trials.
Shown as the dashed curves are the true matrix elements and the true ensemble-averaged taMSD for each state, determined using the known diffusive states of trajectories, while the shaded bands represent their standard deviations.
}
}
\end{figure}

To permit pEMv2 to deal with  tracks containing transitions, we implemented a procedure that splits
long trajectories into shorter trajectories.
For the 120-step data set,
Fig.~\ref{Case2_1}D shows that the log-probability of various model sizes for tracks,
split into  5, 10, 15, 20, or 30 steps,
yields the correct model ($K=3$) for  bin sizes  less than 30 steps, in agreement with Fig.~\ref{Case2_1}A.
 Given that the three state model is correct, Fig.~\ref{Case2_1}E shows the fraction of the total number of steps that are assigned to the correct
diffusive state
 as a function of the bin size. 
 Evidently, this procedure  yields a significant improvement in the fraction of steps correctly assigned
 compared to analysis of the 120-step data set, shown in Fig. \ref{Case2_1}B, presumably as a result of decreasing the
 number of transitions per track from $\sim$1.1 per track for the
 120-step data set to $\sim$0.1 transitions per track for the bin size of 10 steps (Fig. \ref{Case2_1}).
 Moreover, pEMv2 is now able to provide a significant improvement in the quality of the estimates for the covariance elements and taMSD for each diffusive state (Fig. \ref{Case2_2}), as well as reasonable estimates of the transition matrix (Fig. S8).

Is there an optimal bin size? Indeed, Fig.~Fig. \ref{Case2_1}E shows that the fraction of steps correctly assigned exhibits a maximum at a bin size of 10 steps and decreases for smaller and larger bin sizes.
It turns out that using smaller bin sizes may render the results of pEMv2 more susceptible to misclassification (Fig. S9).
Since information of confinement manifests as anti-covariances between neighboring displacements each time a particle ``bounces'' off of the confinement barrier -- if the bin size is too small, then this information is only contained in the few bins which capture such a ``bouncing" event, while other bins would follow an apparent normal diffusion. On the other hand, although including more steps in the bin size allow for more anti-covariance ``bouncing" events, a large bin size also has the undesirable effect of increasing the number of transitions per track,
which can also lead to poorer performance.
Thus,
 the bin size should be chosen to be as large possible, subject to the constraint that the mean number of transitions per trajectory should not be too large.
In this example of case 3, satisfactory results
are obtained by using a level of binning that yields an average of 0.2 transitions per trajectory.

\subsubsection{\label{transitions}Determining the optimal bin size}

\begin{figure}[!htbp]
\centering{\includegraphics[width = 3.35in]{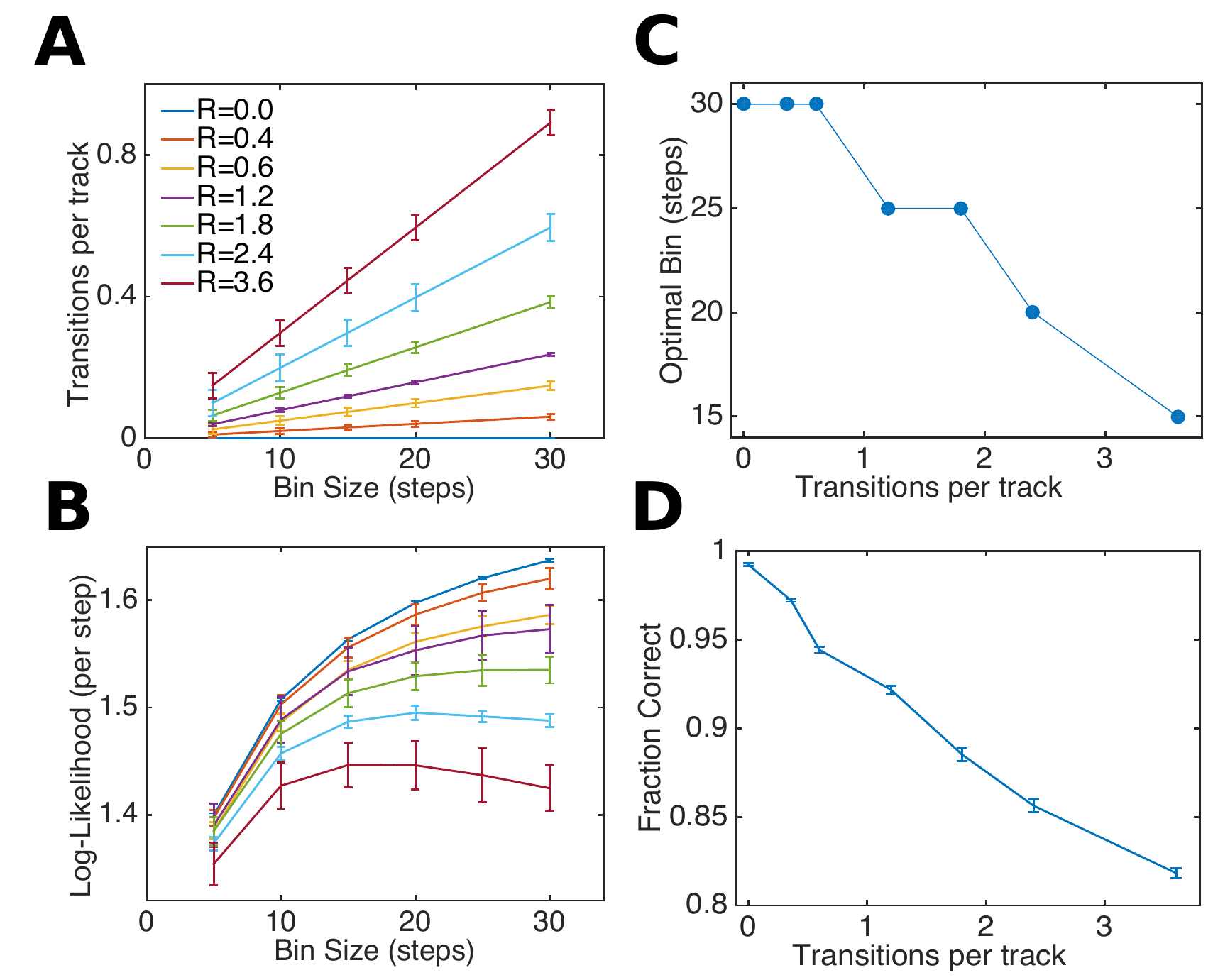}}
\caption{\label{OptimalBin} \small {
Dependence of the optimal bin size on transition rate.  (A) Average number of transitions per track versus bin size for various transition rates, $R = \{0, ~0.36, ~0.6, ~1.2, ~1.8, ~2.4, ~3.6 \}$ transitions per track (each shown in a different color). (B) Average log-likelihood value  per step versus bin size for various transition rates. 
(C) Optimal bin size versus the mean number of transitions per track, determined by the maximum log-likelihood per per step.  (D) Average fraction that the classified diffusive state matches the simulated diffusive state as a function of the mean number of transitions per track.
 Each error bar represents the observed standard deviation across 5 different sets of simulated particle tracks.  
}
}
\end{figure}

To further elucidate how  pEMv2's performance depends on the level of transitions and how to determine the optimal bin size in an unsupervised manner,
we generated a number of data sets containing 3,000 synthetic particle tracks with diffusive states given according to case 4 (Table \ref{Table2}),
and
 with varying mean numbers of transitions per track
 ($R = \{0, ~0.36, ~0.6, ~1.2, ~1.8, ~2.4, ~3.6 \}$ transitions per track).
All of the track lengths were constant with $N$ = 120 steps.
 For each data set, we applied pEMv2 with bin sizes ranging from 5  to 30 steps.
 The mean number of transitions per trajectory for
 each bin size is shown as a function of bin size in Figure \ref{OptimalBin}A.

By applying pEMv2 to each of these data sets, the BIC's log-probability found the correct model size ($K=4$) 
when the transition rates were low ($R < 0.6$ transitions per track),
 irrespective of the bin size used (Fig. S10).
Even when the transition rates increase  ($R = 2.4$ and $R = 3.6$ transitions per track),
the BIC  continues to favor the correct four diffusive state model for smaller bin sizes.
 However, 
the BIC  favors an  incorrect five-diffusive-state model when analyzing data that uses bins containing 30 steps.

Assuming the correct model size ($K=4$), figure \ref{OptimalBin}B shows the average maximum likelihood values as a function of the bin size for each data set. When transition rates are low, the average log-likelihood per step  increases monotonically with bin size, suggesting that in these cases the optimal bin size is larger than the maximum binning used.
For larger numbers of transitions per track, however, a maximum log-likelihood per step is observed within the range of bin sizes examined.  Figure \ref{OptimalBin}C shows that the optimal bin size, determined as the maximum log-likelihood per step from Fig. \ref{OptimalBin}B, decreases as the number of transitions per track increases. 
Not surprisingly, the more transitions that are present, the smaller the bin size should be.
Figures S12 and S13 shows the pEMv2 classification of representative trajectories of the $R=3.6$ data set for various bin sizes. When the bin size is five steps, spurious transitions are frequently found,
which we ascribe to the relatively larger statistical fluctuations that necessarily accompany smaller bin sizes.
 As the bin size becomes larger, statistical fluctuations are reduced.
However, if the bin size becomes too large ($B=30$),
 the corresponding higher rate of transitions per track limits pEMv2's ability to classify diffusive states accurately.
 Evidently, the optimal bin size balances the accuracy of the covariance matrix elements,
 which becomes better-determined with larger bin sizes, against the number of transitions per track,
 which mix
the covariance matrix elements of different diffusive states,
leading to poorer  pEMv2 performance.

Although pEMv2, using the optimal bin size, is able to uncover the correct numbers of diffusive states and characterize each diffusive state reliably (Fig. S11), the overall accuracy of pEMv2's classification decreases as the number of transitions
per track increases,
as is indicated by the fraction of correctly classified steps, plotted in Fig. \ref{OptimalBin}D.
 Even though the optimal bin size lowers the \emph{effective}
number of transitions per track, the decreased performance may be due to the higher \emph{absolute}
number of transitions for the data sets with higher R (Fig. \ref{OptimalBin}A).   
Notwithstanding, the ensemble behavior of each diffusive state can still be captured accurately when the optimal bin size determined by the maximum likelihood per displacement is used (Fig. S11).

\section{Conclusions}

In this paper, we introduced the likelihood functions for two canonical modes of non-normal diffusion, namely confined diffusion, and fractional Brownian motion.  We showed that the maximum likelihood estimates provide a significant improvement in comparison with traditional MSD analysis.  We introduced a model selection scheme, namely mleBIC, to determine the underlying diffusion model that best represents the motions of a diffusing particle. We demonstrated that while mleBIC is quite successful at classifying tracks without localization noise; classification of tracks with localization noise was limited, especially for short trajectories.
Although, in this paper we restricted consideration to
particles undergoing
 normal diffusion, 
confined diffusion,  and fBm and immobile particles,
extensions to other diffusion models
can be added facilely by incorporating these models into mleBIC,
once the likelihood functions are known. 

To take advantage of a systems-level approach, we introduced an updated version of pEM analysis, namely pEMv2, 
that determines the number of unique covariance structures within a population of particle trajectories,
 thereby bolstering the statistics of individual trajectories.
A key output from the pEMv2 algorithm is the posterior probability, $\gamma_{mk}$, that particle trajectory $m$ realizes diffusive state $k$. For the selected model,
one simple and useful way  to categorize a particular trajectory to a particular diffusive state
is to assign the trajectory to the diffusive state that realizes the largest posterior probability,
as in Fig.~\ref{Case1_4}.

 When analyzing simulated trajectories that transition between different normal/non-normal diffusive states, pEMv2 was able to determine the covariance structure of each diffusive state quite reliably. We also demonstrated the rationales for the selection of the free parameters in pEMv2, which includes the number of covariance features and the bin size.  The number of off-diagonal covariance matrix elements to include can be set to the value for which the observed ensemble-averaged covariance matrix element have just decayed to zero, thereby only including informative covariance terms in the analysis. 
 We have shown that
an optimal bin size may be determined by rerunning pEMv2 for various bin sizes, and selecting the bin size that yields the highest likelihood for a given model size.
In practice, because the model size is unknown {\em a priori} for experimental data, we envision running pEMv2
for different model sizes and different bin sizes to find these conditions.
  Importantly, pEMv2 is rooted in physical principles of stochastic processes.
 Applying non-physical clustering methods to the same data, such as k-means clustering, lead to poor characterization of the underlying diffusive states \cite{Koo}. 

Since pEMv2 does not make any intrinsic assumptions of the underlying diffusion model, besides that it follows a Gaussian process, pEMv2 is essentially a diffusion-model-free approach. Characterization of each covariance structure to determine the diffusion mode and properties can then be performed {\em post-hoc}.
Specifically, traditional analyses can then be applied for each diffusive state, such as calculation of the ensemble-average taMSD and the ensemble-average velocity autocorrelation function.
Such a procedure provides a more reliable representation of the diffusive behavior compared to individual trajectories, which suffer from limited statistics.  

One drawback to pEMv2, is that information of the diffusive state across every bin is treated independently.  Thus, when the bin size becomes small, spurious states may occur more frequently.
A key benefit of a hidden Markov model (HMM) approach is that spurious transitions can
be intrinsically penalized by maximizing a likelihood function which includes a transition matrix.
We envisage that, in the future, pEMv2 can be extended to a HMM of multivariate Gaussians, to mitigate spurious transitions.  In turn, this approach will improve the temporal resolution by making it possible
to reduce the bin sizes.
However, if transition rates are inhomogeneous across the cell,
any HMM approach that assumes a single transition matrix for the entire cell,
would not be able to properly capture that inhomogeneity.
Notwithstanding, current HMMs applied to SPT data, namely vbSPT and HMM-SPT \cite{monnier2015inferring}, apply a HMM of univariate Gaussians, which is equivalent to using a bin size of 2 steps ($f=0$) and thus only using information of the first covariance term.  Thus, these HMM analyses overlook localization noise, which introduces correlations between nearest-neighbor displacements, rendering each displacement non-Markovian.  Moreover, neither method can properly account for non-normal diffusion models such as confined diffusion and fBM.  

Unfortunately, there is no strict rule
concerning how many tracks are needed for pEMv2 to return accurate results.
Rather, the amount of data needed depends on the complexity of the diffusive states involved, as discussed previously \cite{Koo}. In practice, we recommend that pEMv2 users complement their SPT analysis of experimental data with an analogous analysis of simulated tracks that recapitulate the diffusive complexity determined by pEMv2.
In this way, the user can determine the reliability of pEMv2 for the data in hand, and thereby gain confidence in the results provided by pEMv2. 

With the ability to handle normal/non-normal diffusive states which contain transitions between different diffusive states, we envision pEMv2 can help to uncover more accurate information regarding the diffusive states which occur inside live cells with single molecule resolution.  This analysis sets the benchmark for all future single particle tracking analysis, to begin to understand the spatio-temporal biochemistry of diffusing particles inside live cells with single molecule resolution.

\section{\label{methods}Methods}	

\subsection{Simulation procedure}	

Synthetic particle trajectories undergoing normal diffusion are generated using the recursion given by Eq. \ref{eqn:stochastic1}, with $\Sigma_{i,j} = 2D\Delta t \delta_{i,j}$ and $x_0=0$.  

To generate synthetic particle trajectories undergoing normal diffusion confined in a finite square geometry with size $-L$ to $L$, we simulate displacements that follow normal diffusion.  At each time step, if the new position falls outside of the finite domain, then the simulated position is set such that the difference between the proposed position and the boundary is reflected, \textit{i.e.} Neumann boundary condition, but the total distance traveled remains the same as if the wall were not present. Here, the starting position of each trajectory is at the center of the confinement boundary.

Synthetic particle trajectories undergoing fractional Brownian motion are generated using the recursion: given by Eq. \ref{eqn:stochastic1} with $v=0$ and $\Sigma_{i,j}$ given by Appendix \ref{covariance}.  Here, the square root of the covariance matrix is determined with the Cholesky decomposition, {\it i.e} $\Sigma = LL^T$, where $L$ is the Cholesky lower triangular matrix.  We then generate a vector of normally distributed random numbers $\mathbf{W} = \{W_d\}_{d=1}^{D}$, where $D$ is the number of displacements of the particle trajectory, and apply a matrix multiplication according to $\Delta \mathbf{x} = L\mathbf{W}$. 
The positions are then reconstructed by calculating the cumulative sum of the displacements $x_i = x_0 + \sum_{j=1}^i \Delta \mathbf{x}_j$, with $x_1 = 0$.  For each particle trajectory, the process is carried out separately for two spatial dimension and are then combined to form the true two-dimensional (2D) positions of the synthetic particle trajectory. 

To incorporate transitions between diffusive states within each trajectory, we first generated a random Markov chain, with a known transition matrix, $A$, to specify the state sequence of each particle track displacements. For each state, the displacements are simulated according to the properties of the diffusive state.  The particle trajectories are then reconstructed their positions by calculating the cumulative sum of the displacements  $x_i = x_0 + \sum_{j=1}^i \Delta \mathbf{x}_j$, with $x_1 = 0$.  Each time the Markov state goes to a confined diffusion state, the confinement boundaries are reset with the initial position at the center.  When the Markov state switches to another diffusive state, information of the confinement boundaries is forgotten.  

Dynamic localization noise is incorporated into the positions by simulating 32 micro-step displacements ($\delta t = \Delta t/32$) time steps and averaging 32 successive positions. The net effect is an exposure time equal to the frame duration of 32~ms. Static localization noise is included by adding a normally distributed random number with zero mean and variance, $\sigma_{sim}^2$, to each motion-blurred position. To generate a collection of particle trajectories, the population fractions are used to determine the number of particle trajectories that are initialized to each diffusive state.  Here, population fractions serve as the percentage that the initial state of each trajectory begins with.  

\subsection{MSD analysis}

For a stationary sequence of $T$ 2D particle positions, $\mathbf{x} = \{x(t), y(t)\}$ for $t=1$ through $T$, each separated one from the next by a time, $\Delta t$, the taMSD is given according to \cite{qian1991single,kusumi1993confined}:
\begin{eqnarray*}
	\overline{\delta(\Delta_n,T)} &=& \frac{1}{T-\Delta_n}\sum_{t=1}^{T-\Delta_n}
					\left( x(t+\Delta_n)-x(t)\right)^2 \\
					& & + \left( y(t+\Delta_n)-y(t)\right)^2
	\label{eq:taMSD}
\end{eqnarray*}
where $\overline{\delta(\Delta_n,T)}$ is the taMSD for the $n$th time lag, $\Delta_n = n\Delta t$ and the bar on top of $\delta(\Delta_n,T)$ is used to distinguish the time average.

We generate the taMSD for the first 14 time lags and employ an unweighted non-linear least squares fit with diffusion models given in Table III,
where $\sigma_0$ represents the static localization noise and $\sigma$ represents the combined static and dynamic localization noise terms.

\begin{table}[!htbp]
\label{table:msd}
\small
\centering{
\begin{tabular}{|c|c|}
\hline 
Mode & MSD model\\
\hline &\\ 
Normal & $4D\Delta t + 4\sigma_0^2 - \frac{2}{3}D\Delta t$\\  [2ex]
&\\ \hline &\\
Confined & $\frac{L^2}{3} - \frac{32L^2}{\pi^4} \sum_{k=1,odd}^\infty \frac{1}{k^4} \exp\left[-\left( \frac{k\pi}{L} \right)^2 D n \Delta t \right] + 4\sigma^2 $  \\ [2ex]
&\\ \hline &\\
fBm  & $4D\Delta t^\alpha + 4\sigma^2$ \\ [1ex]
&\\ \hline
\end{tabular}
\caption{{MSD models for canonical modes of diffusion.}} 
}
\end{table}

\subsection{\label{mleanalysis}MLE analysis}

For a given diffusion model, the maximum likelihood, or equivalently the minimum negative log-likelihood, is found by employing a constraint,  gradient-based, numerical optimization algorithm in MATLAB (Mathworks), namely {\tt fmincon}.  
At each optimization step, however, the log-likelihood function requires the calculation of the log-determinant and the inverse of the covariance matrix.  When particle tracks are long or the elements of the covariance matrix are very small, the log-determinant of the covariance matrix can run into numerical underflow issues.  To make this optimization procedure more robust, we employ an eigenvalue decomposition of the covariance matrix, $\Sigma = P\Lambda P^{T}$, where $P$ is a matrix of  the eigenvectors with their corresponding eigenvalues given along the diagonal of $\Lambda$.  The log-determinant is given by the product of the eigenvalues or equivalently the sum of the log eigenvalues, {\it i.e.} $\ln \det(\Sigma) = \ln \prod_i^D \lambda_i = \sum_i^D \ln \lambda_i$.  The inverse is given by $\Sigma^{-1} = P\Lambda^{-1}P^T$.  

In summary, for a given single particle trajectory, maximum likelihood estimation yields the parameter estimates, log-likelihood value, and Hessian for each candidate diffusion model. From this information, the BIC (Eq. \ref{eqn:BIC}) can be calculated for each diffusion model.  Once the BIC for each diffusion model has been calculated, the model probability for each diffusion model can be calculated according to Eq. \ref{eqn:BICprob}. Classification is determined by the diffusion model which yields the highest model probability.  

\subsection{pEMv2 analysis}

pEMv2 analysis was performed with the Matlab script provided in the Supplemental Materials.  Briefly, our pEMv2 procedure employs the EM algorithm on the original set of particle trajectories with random initial parameter values.  pEMv2 then reemploys the EM on a Monte Carlo bootstrap set of the original particle trajectories, which serves to perturb the likelihood surface with the aim  that a local maximum may no longer be a maximum in the perturbed likelihood surface.  Upon completion of a perturbation trial, we verify whether a  higher likelihood has truly been found by calculating the likelihood using the pEMv2-converged parameters with the original dataset. If the pEMv2-converged parameters indeed yield a higher likelihood, then the EM parameters are updated by reemploying the EM algorithm initialized with the new pEMv2 parameter estimates on the original dataset. Otherwise, the pEMv2 estimates remain unchanged. This process is repeated until a predetermined number of perturbations have been executed and yield no advance. 

To generate each set of random initial values, $K$ random numbers between 0 and 1 are drawn from a uniform distribution.  The initial population fractions, $\{\pi_k^0\}_{k=1}^K$, are given by normalizing these random numbers so that the sum is equal to 1. The first covariance values are set using the initial randomly-chosen population fractions and the empirical cumulative covariance distribution function.  By dividing the cumulative distribution function into K regions proportional to the initial population fractions, the initial covariance value of diffusive state k is then picked as the diffusivity corresponding to the midpoint of region k of the cumulative probability distribution, namely to $\sum_{j=1}^{k-1} \pi_j^0 +\frac{\pi_k^0}{2}$.  Particle tracks are then classified to each diffusive state by their distance to the first covariance  values.  The remaining covariance values for each diffusive state is selected by averaging the classified covariance terms. Thus, we achieve an initialization that serves as a non-parametric method to randomly sample from the observed distribution of diffusion coefficients.  We found this method produces better random initializations than a k-means clustering over the whole covariance matrix. In practice, we found k-means clustering converges to similar values over a wide range of parameter space. In addition, k-means tends to weed out diffusive states with low population fractions, when two diffusive states are close in proximity.

To improve pEMv2's performance, we applied 5 random initialization trials of the EM and used the parameters of the trial which yielded the highest likelihood value.  We then applied 100 perturbation trials.  This was done for each diffusive state starting from $K=1$ and incrementing $K$ till pEMv2 finds a lower BIC value.  Upon completion of pEMv2, the returned parameters include the population fractions and covariance matrices of each diffusive state, along with the posterior probabilities of each particle trajectory. 

pEMv2 is written in MATLAB (Mathworks) and is freely available at https://GitHub.com/MochrieLab/pEMv2.

\newpage

\appendix
\section{\label{covariance}Covariance matrix for particle track displacements without localization noise}

\begin{table*}[!htbp]
\label{table:covariance}
\caption{Analytical covariance matrix of particle track displacements separated in time by $\Delta t$ for canonical diffusion modes, namely normal diffusion, confined diffusion, and fractional Brownian motion, with $D$ diffusion coefficient, $L$ confinement size, and $\alpha$ anomalous exponent. The hat, $\mathbf{\tilde{\Sigma}}$, represents the covariance matrix without localization error corrections. \\
}  
\small
\centering{
\begin{tabular}{|c|c|}
\hline
Mode & Covariance matrix ($\mu\rm{m}^2\rm{s}^{-1}$)\\
\hline &  \\ 
Normal & $\mathbf{\tilde{\Sigma}}^{normal}({i,j}) =  2D\Delta t \delta_{i,j}$\\  [2ex]
\hline & \\  
Confined & $
\mathbf{\tilde{\Sigma}}(i,j)^{confined} 
	 = \left\{  
        \begin{array}{ll}
             \frac{L^2}{6} - \frac{16L^2}{\pi^4} \sum_{k=1,odd}^\infty \frac{1}{k^4} \Phi(1)  & \quad ,~j=i \\
             \frac{-L^2}{12} + \frac{8L^2}{\pi^4} \sum_{k=1,odd}^\infty \frac{1}{k^4}  \Phi(1) \left(2-\Phi(1) \right) & \quad ,~j= i\pm 1\\
             \frac{8}{\pi^4} \sum_{k=1,odd}^\infty \frac{1}{k^4}\left( -2\Phi({j-i+1}) + \Phi({j-i})+ \Phi({j-i+2}) \right) & \quad ,~{\rm otherwise}
        \end{array}       
        \right.
$\\ 
& where $\Phi_n = \exp\left[-\left( \frac{k\pi}{L} \right)^2 D n \Delta t \right]$.\\ [2ex]
\hline & \\  
fBM 
& $\mathbf{\tilde{\Sigma}}^{fBM}({i,j}) = D\Delta t^\alpha (\left|j-i+1 \right|^\alpha + \left|j-i-1 \right|^\alpha - 2\left|j-i \right|^\alpha)$ \\ [1ex] &\\
\hline
\end{tabular}
}
\end{table*}

\newpage
\section*{Acknowledgments}
This work was supported by NSF PHY 1305509, and by
 the Raymond and Beverly Sackler Institute for Physical and Engineering Biology.

\end{document}